\documentclass[12pt]{article}

\pdfoutput=1

\usepackage[top=80pt,bottom=85pt,left=85pt,right=85pt]{geometry}
\usepackage{amsmath,amssymb,graphicx,float,color,cite,subfigure}
\usepackage{caption}
\usepackage[debug,pageanchor=false]{hyperref}
\definecolor{link}{rgb}{.8,.15,.1}
\hypersetup{colorlinks=true,linkcolor=link,citecolor=link,urlcolor=link,linktocpage}

\usepackage[vcentermath]{youngtab}

\makeatletter
\@addtoreset{equation}{section}
\makeatother

\begin{document}

	\begin{titlepage}
		\begin{flushright}
			{CERN-TH-2019-232}\\
			\end{flushright}
	\begin{center}

	\noindent

	{\Large \bf{On AdS$_7$ stability}}

	\bigskip\medskip

	 Fabio Apruzzi,$^1$ G.~Bruno De Luca,$^2$ Alessandra Gnecchi,$^3$\\ Gabriele Lo Monaco,$^{2,4,5}$ Alessandro Tomasiello$^2$\\   

	\bigskip\medskip
	{\footnotesize 
$^1$ 	Mathematical Institute, University of Oxford, \\
	Andrew-Wiles Building, Woodstock Road, Oxford, OX2 6GG, UK
\\	
	\vspace{.3cm}%{.1cm}
$^2$ Dipartimento di Fisica, Universit\`a di Milano--Bicocca, \\ Piazza della Scienza 3, I-20126 Milano, Italy \\ and \\ INFN, sezione di Milano--Bicocca
\\	
	\vspace{.3cm}%{.1cm}
$^3$ Theoretical Physics Department, CERN, Geneva, Switzerland
\\	
	\vspace{.3cm}%{.1cm}
$^4$ 	Department of Physics, Stockholm University, AlbaNova, 10691 Stockholm, Sweden 		
\\	
	\vspace{.3cm}%{.1cm}
$^5$ 	Institut de Physique Th\'eorique, CEA Saclay, 91191 Gif-sur-Yvette Cedex, France   
		}

	\vskip .5cm %.3cm
	{\small \tt Fabio.Apruzzi@maths.ox.ac.uk, g.deluca8@campus.unimib.it, alessandra.gnecchi@cern.ch, g.lomonaco1@campus.unimib.it, alessandro.tomasiello@unimib.it}
	\vskip .9cm %.6cm
	     	{\bf Abstract }
	\vskip .1in
	\end{center}

	\noindent AdS$_7$ supersymmetric solutions in type IIA have been classified, and they are infinitely many. Moreover, every such solution has a non-supersymmetric sister. In this paper, we study the perturbative and non-perturbative stability of these non-supersymmetric solutions, focusing on cases without orientifolds. Perturbatively, we first look at the KK spectrum of spin-2 excitations. This does not exhibit instabilities, but it does show that there is no separation of scales for either the BPS and the non-BPS case, thus proving for supersymmetric AdS$_7$ a well-known recent conjecture. We then use 7d gauged supergravity and a brane polarization computation to access part of the spectrum of KK scalars. The result signals an instability for all non-supersymmetric solutions except those that have a single D8 on each side. We finally look at non-perturbative instabilities, and find that NS5 bubbles make these remaining solutions decay.

	\noindent

	\vfill
	\eject

	\end{titlepage}

\tableofcontents

\section{Introduction} % (fold)
\label{sec:intro}

Finding effective strategies to break supersymmetry is of paramount importance in string theory. 
While various geometrical techniques have been developed over the years to find vacuum solutions that preserve supersymmetry, there is a relative scarcity of strategies for finding vacua that do not. Even when solutions are found, they are vulnerable in various ways. For example, massless moduli that might be present in the supergravity approximation are not protected against runaway potentials, that would reveal they are not solutions at all in string theory. 

Equally common is the presence of instabilities. This can manifest itself as the presence of an excitation that tends to grow over time. For a Minkowski solution this is a field with negative mass squared $m^2$; for AdS$_d$ solutions, fields with $m^2$ above the so-called Breitenlohner--Freedman (BF) bound $-\frac{(d-1)^2}{4L^2}$ \cite{breitenlohner-freedman} are stabilized by forming a standing wave that reflects off the gravitational potential, but below this bound they grow and destabilize the solution. (At a non-linear level, generic small perturbations seem to form black holes over long enough times \cite{bizon-rostworowski,bizon}.) Another potential form of instability is the formation of bubbles of a vacuum with lower cosmological constant. This process is similar to the tunnel effect in quantum mechanics, and is non-perturbative in $G_\mathrm{N}$. A thin bubble has a non-vanishing probability of being created if it has a charge density larger than its tension.  If this is the case, the original false vacuum is called metastable.

Supersymmetric vacua are believed to be protected against both types of instabilities \cite{breitenlohner-freedman,gibbons-hull-warner,cvetic-griffies-rey}. In absence of supersymmetry, however, the situation is less clear. It is relatively easy to find solutions which are perturbatively stable, i.e.~where all linearized fluctuations have masses above the BF bound. But bubble creation is much harder to exclude: one should estimate the nucleation probability relative to all possible false vacua. This issue is especially important in string theory, with its infamously large set of vacua. 

In view of this, one can wonder whether there are any non-supersymmetric vacua in string theory which are completely stable. A possible line of attack was suggested in \cite{ooguri-vafa-ads,freivogel-kleban} by generalizing the weak gravity conjecture (WGC) \cite{arkanihamed-motl-nicolis-vafa}. The original WGC posits that in all models that include quantum gravity a particle should always exist whose charge is larger than its mass. For extended objects, it is natural then to conjecture that a brane should always exist whose charge density is larger than its tension. Applying this to codimension-one branes implies that there always exists a bubble with non-zero nucleation probability. 

Existing de Sitter models are  indeed metastable; for AdS, the statement is however perhaps a little more surprising, especially from a holographic point of view. Once they get created, bubbles reach the boundary in finite time, gobbling up all of spacetime. Moreover, as pointed out in \cite{ooguri-vafa-ads}, even a small region in the boundary is influenced by an infinite-volume region of AdS; with a finite probability of nucleation, the decay time should be zero. This would seem to indicate that it doesn't make sense to describe a CFT dual of a metastable AdS vacuum. The conjecture of \cite{ooguri-vafa-ads,freivogel-kleban} would then imply that in string theory there is no example of AdS/CFT without supersymmetry. 

It makes sense to test this conjecture with concrete non-supersymmetric AdS solutions. Some examples had already been analyzed in the literature. In \cite{maldacena-michelson-strominger} examples were given of metastable AdS$_3$ solutions. In \cite[Sec.~4.1.2]{gaiotto-t}, a set of non-supersymmetric AdS$_4$ solutions were found to be unstable to decays mediated by D2 bubbles. (In that paper, a slightly different point of view was taken, involving domain walls in Poincar\'e coordinate, but we explain in section \ref{sub:glo} below that the two points of view yield the same results.) \cite{narayan-trivedi} found vacua that are marginally stable for the considered decay channels. More recently, for example \cite{antonelli-basile} considered the decay of models with brane supersymmetry breaking.

In this paper we test the conjecture on a large set of AdS$_7$ solutions. The full set of supersymmetric AdS$_7$ solutions in type II is known \cite{afrt,10letter,cremonesi-t}. We don't have a similar classification for all solutions, but a consistent truncation technique \cite{passias-rota-t} allows us to say that every supersymmetric solution has a supersymmetry-breaking sister solution. 

We will analyze the perturbative and non-perturbative stability of these vacua. Establishing the presence of perturbations below the BF bound would require a Kaluza--Klein (KK) reduction, which is rather challenging, especially in the presence of non-constant warping function and dilaton, which these solutions have. Only the spin-2 tower is easy, thanks to a result of \cite{bachas-estes} that gives the mass operator as a certain Laplacian modified by the warping function. This was used in \cite{passias-t} to find the spin-2 spectrum of both the supersymmetric vacua and their non-supersymmetric sisters. Those results do not exhibit any instability; here we also point out that they establish for AdS$_7$ ${\mathcal N}=(1,0)$ another recent conjecture, namely the absence of scale separation \cite{lust-palti-vafa}.

For the spectrum of spin-0 excitations, we instead use a shortcut. In \cite{deluca-gnecchi-lomonaco-t}, a seven-dimensional gauged supergravity was found, that describes simultaneously all the AdS$_7$ solutions, both with and without supersymmetry, with a fixed total D6-brane charge. While this is not expected to be a consistent truncation, it reproduces some of the results expected in a full ten-dimensional analysis. The masses of the scalars in this 7d theory then provide a proxy for some of the scalars in the KK tower.\footnote{This approach was already used for these vacua in \cite{apruzzi-dibitetto-tizzano}; the gauged supergravity used there, while being more solid, only had three scalars, none of which destabilized the vacua.} In fact we are also able to reproduce this computation from a non-abelian potential in the spirit of \cite{myers}, interpreting the D8-branes in the AdS$_7$ solutions as polarized D6s. 

We conclude from this perturbative analysis that there is an instability for almost all solutions; it tends to polarize D6-branes, and to reduce the number of D8-branes.  The mass of the offending tachyon is $m^2=-12/L^2$ for all solutions. While our methods are more solid when the D8-branes are small, this mass is well below the BF bound, and that seems to suggest our result is valid in general. The instability leads to the maximally polarized vacua, that only have two D8-branes, separating a region where the Romans mass $F_0=0$. For solutions with D6-branes, our polarization instability competes with one found in \cite{danielsson-dibitetto-vargas-swamp}, which seems to spread D6s over the internal space and whose endpoint is presumably a solution of \cite{blaback-danielsson-junghans-vanriet-wrase-zagermann-1,blaback-danielsson-junghans-vanriet-wrase-zagermann-2} with smeared D6s. 

Turning to non-perturbative stability, we find that an NS5-brane bubble can get nucleated. The ratio of the bubble charge density to tension is again the same for all solutions, and is larger than one, allowing for tunneling effects to solutions with a lower amount of NSNS flux. This process can then repeat itself. As usual, when $N$ becomes too small the supergravity approximation breaks down. In any case, taken together, these results seem to point to an instability for all these solutions, as long as they are under control.

We begin section \ref{sec:ads} with a quick review of AdS$_7$ solutions, with an emphasis on non-supersymmetric ones; we also review the gauged supergravity of \cite{deluca-gnecchi-lomonaco-t} and the spin-2 computation of \cite{passias-t}, making it clear that it applies to all solutions and proving scale separation. In section \ref{sec:pert} we discuss perturbative instabilities, using both gauged supergravity and a brane polarization computation as a guide. In section \ref{sec:bubbles} we consider non-perturbative instabilities; after a brief review of various approaches, we show that NS5-brane bubbles can be created. Finally in section \ref{sec:dw} we consider flat domain walls. These represent holographic RG-flows; we show two BPS classes of examples, which correspond to two ways of Higgsing the dual SCFTs.

% section intro (end)

\section{AdS$_7$ review} % (fold)
\label{sec:ads}

Supersymmetric AdS$_7$ solutions were classified in a series of papers \cite{afrt,gaiotto-t-6d,10letter,cremonesi-t}. While there are no such solutions in IIB, there are infinitely many in IIA supergravity. The fields in every solution have the same functional dependence on a piecewise-cubic real function $\alpha=\alpha(z)$ of a single variable $z$ belonging to an interval $[0,N]$. The internal space $M_3$ always have the topology of an $S^3$. Several types of internal sources are possible: D6- and D8-branes, O6- and O8-planes. In this paper we are going to mostly focus on solutions without O-planes. 

All these supersymmetric solutions have a non-supersymmetric counterpart; we will review them in section \ref{sub:sol}, along with the consistent truncation originally used to find them. Since in later sections we will be interested in the spectrum of perturbations, in section \ref{sub:scale} we will review what is currently known about the tower of spin-2 excitations around both the supersymmetric and non-supersymmetric vacua, adapting the results of \cite{passias-t} to the more convenient set of coordinates found in \cite{cremonesi-t} and used in this paper. In particular this shows that there is no scale separation for supersymmetric AdS$_7$ solutions, even if in some cases some reasonable-looking estimates might lead one to think otherwise. Finally in section \ref{sub:gsugra} we will review the gauged supergravity approach of \cite{deluca-gnecchi-lomonaco-t}, which we will use in section \ref{sec:pert} to analyze part of the perturbative spectrum.

\subsection{AdS$_7$, with and without supersymmetry} % (fold)
\label{sub:sol}

In \cite{passias-rota-t} a consistent truncation was found, from any of the supersymmetric AdS$_7$ solutions to minimal gauged supergravity in seven dimensions. Its fields are
\begin{equation}\label{eq:gra-mult}
\left\{ g_{\mu \nu}, \psi^A_{\mu}, A^i_{\mu}, \chi^A, B_{\mu\nu}, X \right\}\,, 
\end{equation}
where the indices $A=1,2$ and $i=1,2,3$ are R-symmetry indices. The gravitini $\psi^A_\mu$ and the dilatini $\chi^A$ transform in the $\mathbf{2}$ of $SU(2)_R$ while the three vectors $A^i_\mu$ are valued in the Lie algebra of $\mathrm{SU}(2)$, and can be thought as the generators of R-symmetry. 

The truncation to this theory was nicely reexpressed in \cite{malek-samtleben-vallcamell} in the more convenient coordinates introduced in \cite{cremonesi-t}.\footnote{For the scalar $X$ we follow however the convention in \cite{passias-rota-t} rather than that in \cite{malek-samtleben-vallcamell}.}  The uplift depends on a single piecewise-cubic function $\alpha(z)$ of a single variable $z \in [0,N]$; it should satisfy
\begin{equation}\label{eq:ddd}
	\dddot \alpha = - 162 \pi^3 F_0 \,,
\end{equation}
where $\dot{(\ )}\equiv \partial_z ()$. $F_0$ is the Romans mass parameter, which is quantized according to $F_0 = \frac{n_0}{2\pi}$. The integer $n_0$ is allowed to jump at integer values of $z$. Flux quantization also requires that $\frac{\ddot\alpha(i)}{81 \pi^2} \in \mathbb{Z}$ for every $i\in \mathbb{Z}$.

The uplift formulas are then as follows. The metric reads
\begin{equation}\label{eq:metX}
	\frac1{\pi\sqrt2} ds^2= \frac{120 X^{-1/2}}{V_0} \sqrt{-\frac \alpha{\ddot \alpha}} ds^2_{7}+  X^{5/2}\sqrt{-\frac {\ddot \alpha}\alpha} \left(dz^2 + \frac{\alpha^2}{\dot \alpha^2 - 2 X^5 \alpha \ddot \alpha} Ds^2_{S^2}\right)\,.
\end{equation}
Here 
\begin{equation}\label{eq:Vgs}
	V_0 = X^{-8} (8 X^{10} + 8  X^5 -1)
\end{equation}
is proportional to the potential of minimal 7d gauged supergravity. $D s^2 = Dy^i D y^i$, $D y_i \equiv d y_i + \epsilon_{ijk} A_j y_k$ is the metric of a round $S^2$, expressed in embedding coordinates $y_i$ such that $y_iy_i=1$, and fibred by $A$ over the seven-dimensional spacetime. The dilaton is given by 
\begin{equation}\label{eq:phiX}
 e^\phi=162\cdot 2^{1/4}\pi^{5/2} X^{5/4}\frac{(-\alpha/\ddot \alpha)^{3/4}}{\sqrt{\dot \alpha^2-2 X^5 \alpha \ddot \alpha}}\ .   
\end{equation}
The possible fluxes are $F_0$, $F_2$  and $H=dB$:
\begin{equation}\label{eq:BX}
	B=\pi \left( -z+\frac{\alpha \dot \alpha}{\dot \alpha^2-2 X^5\alpha \ddot \alpha}\right) {\rm vol}_{S^2}\ ,\qquad F_2 =  \left(\frac{\ddot \alpha}{162 \pi^2}+ \frac{\pi F_0\alpha \dot \alpha}{\dot \alpha^2-2 X^5\alpha \ddot \alpha}\right) {\rm vol}_{S^2}\ ,
\end{equation}
where now $\mathrm{vol}_{S^2}\equiv \epsilon_{ijk}y^i Dy^j Dy^k$ is the (fibred) volume form of the $S^2$.

As we can see from (\ref{eq:Vgs}), the theory has two vacua:
\begin{equation}\label{eq:X-vac}
	X=1 \, ,\qquad X=2^{-1/5}
\end{equation}
 The $X=1$ vacuum is supersymmetric; applying the uplift formulas (\ref{eq:metX})--(\ref{eq:BX}) to it we obtain the solutions of \cite{afrt,gaiotto-t-6d,10letter,cremonesi-t}. The $X=2^{-1/5}$ vacuum is non-supersymmetric. Applying to it the uplift formulas gives another solution of the ten-dimensional equation of motion. Therefore every AdS$_7$ supersymmetric solution has a supersymmetry-breaking sister. 

For readability and completeness let us give here both explicit solutions obtained from (\ref{eq:metX})--(\ref{eq:BX}). We take the radius of the AdS$_7$ metric to be $L_\mathrm{AdS_7}=1$. In the supersymmetric case $X=1$, the metric and fields read 
\begin{subequations}\label{eq:susy}
\begin{align}
	&\frac1{\pi} ds^2= 8\sqrt2\sqrt{-\frac \alpha{\ddot \alpha}}ds^2_{{\rm AdS}_7}+ \sqrt2\sqrt{-\frac {\ddot \alpha}\alpha} \left(dz^2 + \frac{\alpha^2}{\dot \alpha^2 - 2 \alpha \ddot \alpha} ds^2_{S^2}\right)  \,;\\
	&e^\phi=162 \cdot 2^{1/4}\pi^{5/2}  \frac{(-\alpha/\ddot \alpha)^{3/4}}{\sqrt{\dot \alpha^2-2 \alpha \ddot \alpha}}\ ; \qquad \qquad \qquad
	 \text{(supersymmetric)}\\
	&B=\pi \left( -z+\frac{\alpha \dot \alpha}{\dot \alpha^2-2 \alpha \ddot \alpha}\right) {\rm vol}_{S^2}\ ,\qquad F_2 =  \left(\frac{\ddot \alpha}{162 \pi^2}+ \frac{\pi F_0\alpha \dot \alpha}{\dot \alpha^2-2 \alpha \ddot \alpha}\right) {\rm vol}_{S^2}\ .
\end{align}
\end{subequations}	
For the non-supersymmetric case $X^5=\frac12$, 
\begin{subequations}\label{eq:nonsusy}
\begin{align}
	&\frac1{\pi} ds^2= 12\sqrt{-\frac \alpha{\ddot \alpha}}ds^2_{{\rm AdS}_7}+ \sqrt{-\frac {\ddot \alpha}\alpha} \left(dz^2 + \frac{\alpha^2}{\dot \alpha^2 -  \alpha \ddot \alpha} ds^2_{S^2}\right)  \,;\\
	&e^\phi=162 \pi^{5/2} \frac{(-\alpha/\ddot \alpha)^{3/4}}{\sqrt{\dot \alpha^2- \alpha \ddot \alpha}}\ ; \qquad \qquad \qquad   \text{(non-supersymmetric)}\\
	&B=\pi \left( -z+\frac{\alpha \dot \alpha}{\dot \alpha^2- \alpha \ddot \alpha}\right) {\rm vol}_{S^2}\ ,\qquad F_2 =  \left(\frac{\ddot \alpha}{162 \pi^2}+ \frac{\pi F_0\alpha \dot \alpha}{\dot \alpha^2- \alpha \ddot \alpha}\right) {\rm vol}_{S^2}\ .
\end{align}
\end{subequations}
In both cases, the $\mathrm{SU}(2)$ gauge field has been set to zero, so that now simply $ds^2_{S^2}= d\theta^2 + \sin^2 \theta d \phi^2$ appears.

It is easy to show from the flux formulas above that $\int H= -4\pi^2 N$ and that the integral of $\tilde F_2 = F_2 - B F_0$ over the $S^2$ in the interval $z=(i,i+1)$ is $\int_{S^2} \tilde F_2= -\frac{\ddot\alpha(i)}{81 \pi^2}$; as promised, these integrals are integers by our assumptions below (\ref{eq:ddd}). It should be noticed, however, that the periods of $\tilde F_2$, while quantized, are not gauge-invariant. Large gauge transformations of the NSNS potential are given by $B\to B+ \lambda_2$, where $\lambda_2$ has integral periods; these are represented here by translating $z$ by an integer, and make $\tilde F_2$ change. These gauge transformations are needed in our solutions because the NSNS potential given above has the limit $B\to \pi N \mathrm{vol}_{S^2}$ as $z\to N$; taken literally this would mean that at this point $B$ is not regular, and $H$ has a delta. A large gauge transformation with can however turn this $B$ to a form that goes to zero as $z\to N$. We can place this gauge transformation wherever we like; two popular choices are immediately before $z=N$, or in the region where $F_0=0$, if it exists.

In both the supersymmetric and non-supersymmetric case, the solution has back-reacting branes:
\begin{itemize}
	\item D6-branes at the endpoints $z=0$ or $z=N$. They appear with the boundary condition $\alpha=0$. The number of D6-branes is given by $-\frac{\ddot \alpha}{81 \pi^2}$. If at one of the endpoints there are no D6-branes, then both $\alpha=\ddot \alpha=0$ there.
	\item D8-branes at any $z=k\in \mathbb{Z}$ where $n_0$ changes. Their D6-brane charge is equal to $k$. 
\end{itemize}
O-planes can also be obtained with slightly different boundary conditions, but as we said earlier we are not going to consider them here.

We can then classify the most general solution (without O-planes) in terms of the D-branes it contains. Indeed, $\alpha$ is uniquely determined by the condition that $\alpha(0)=\alpha(N)=0$ and by the piecewise-linear function $\ddot \alpha$. In turn, $\ddot \alpha$ is determined by  its values at the endpoints (which as we saw is proportional to the number of D6-branes) and by its second derivative $\partial_z^4 \alpha $, which is a sum of delta-functions: 
\begin{equation}
	\partial_z^4 \alpha = -\frac1{162 \pi^2}\sum_a f_a \delta(z-a)\,.
\end{equation}
This says there are $f_a$  D8-branes at the positions $z=a$, and with D6-brane charge $a$. 

Often it is useful to group these D8-branes in two sets, according to whether they occur in a region where $F_0=-\frac{\dddot \alpha}{162 \pi^3}$ is positive or negative. If there is a region $[L,R]$ where $F_0=0$, we define $f^\mathrm{L}_a \equiv f_a$ for $a\le L$, and $ f^\mathrm{R}_a \equiv  f_{N_a}$ for $a\ge R$. In this case $\sum_a a f^\mathrm{L}_a= \sum_a a f^\mathrm{R}_a\equiv k$. If $L=R$, one can formally split the D8-branes at $z=L=R$ as a sum of $f^\mathrm{L}_L+ f^\mathrm{R}_R$, in such a way that $\sum_a a f^\mathrm{L}_a= \sum_a a f^\mathrm{R}_a$ still holds. 

Another point of view comes from considering the graph of the function $F_0=-\frac{\dddot \alpha}{162 \pi^3}$ directly. Dividing the graph in the parts below and above zero defines two Young diagrams $\mu^\mathrm{L}$, $\mu^\mathrm{R}$. These are related to the $f$ above by $f^\mathrm{L,R}_a = (\mu^\mathrm{L,R})^t_a- (\mu^\mathrm{L,R})^t_{a+1}$, where ${}^t$ denotes Young diagram transposition. For more details and examples, see \cite{cremonesi-t,deluca-gnecchi-lomonaco-t}.

Let us now see some examples of $\alpha$, describing the corresponding physics. 
\begin{itemize}
	\item The easiest solution is  
	\begin{equation}\label{eq:m0}
		\alpha = \frac{81}2 \pi^2 k z (N-z)\,.  
	\end{equation}
	Here from (\ref{eq:ddd}) we see that $F_0=0$ everywhere. There are $k$ D6-branes at $z=0$, and $k$ anti-D6 at $z=k$.
	\item Perhaps the next easiest solution is 
	\begin{equation}\label{eq:1D6}
		\alpha = \frac{27}2 \pi^2 n_0 z (N^2-z^2)\,.
	\end{equation}
	Here $F_0=\frac{n_0}{2\pi}$ is non-zero and constant everywhere, so there are no D8-branes; there are no D6-branes at $z=0$, while there are $k= n_0 N$ anti-D6 at $z=N$.
	\item If we take
	\begin{equation}\label{eq:2D6}
		\alpha = 81 \pi^2 z (N-z) \left(\frac {k_1}2 + \frac{n_0}6(N+z)\right)\,
	\end{equation}
	there are now $k_1$ D6-branes on the left, and $k_2= k_1+n_0 N$ on the right. Again there are no D8-branes. We see that (\ref{eq:2D6}) reduces to (\ref{eq:1D6}) for $k_1=0$, and to (\ref{eq:m0}) for $n_0=0$.
	\item To introduce D8-branes, we need to write distinct expressions for each of the regions where $n_0$ is constant. For example
	\begin{equation}\label{eq:2D8s}
		\alpha = \frac{27}2 \pi^2 n \left\{ \begin{array}{cc}
			z(3 \mu (N-\mu) - z^2)  & z\in [0,\mu]\,,\\
			\mu(3z(N-z)-\mu^2)  &  z\in [\mu,N-\mu]\,,\\
			(N-z) (3 \mu (N-\mu) - (N-z)^2)\, & z \in [N-\mu,N]
		\end{array}\right.
	\end{equation}
describes a solution with $n_0=\{n,0,-n\}$ in the three intervals of definition. Correspondingly, there are $n$ D8-branes at $z=\mu$ and $n$ D8-branes at $z=N-\mu$. Their D6-charges are given by their positions, so $\mu$ and $N-\mu$ respectively. (If we place the NSNS large gauge transformations in the massless region, the latter becomes $-\mu$.)
\end{itemize}

In this subsection we reviewed how the minimal gauged supergravity approach shows the existence of non-supersymmetric solutions. This theory is useful also for other applications. In \cite{apruzzi-dibitetto-tizzano}, it was used to study an RG-flow among the two solutions (originally found in \cite{campos-ferretti-larsson-martelli-nilsson} in seven dimensions). In \cite{bobev-dibitetto-gautason-trujien} it was used to study brane solutions that reduce to AdS$_7$ in their near-horizon region. In \cite{passias-rota-t} it was pointed out that it implies the existence of RG flows relating the AdS$_7$ solutions to the AdS$_5$ and AdS$_4$ solutions of \cite{afpt,rota-t}.

% subsection sol (end)

\subsection{Absence of scale separation} % (fold)
\label{sub:scale} 

We will now start looking at the spectrum of perturbations around the solutions we just reviewed. 
In particular, in this section we will consider the tower of spin-2 excitations, adapting a computation originally carried out \cite{passias-t}
so that it becomes more obvious that it applies to all solutions --- something perhaps not entirely clear from that reference.

At first, one might attempt to analyze scale separation using some estimates on the scales of KK modes and of the cosmological constant. In \cite{gautason-schillo-vanriet-williams} it was suggested to estimate the scale of the KK modes from the internal curvature, which is a priori reasonably well-motivated from various theorems on spectral operators. This led to the criterion 
\begin{equation}\label{eq:MKK-est}
	\left|\frac{\int_{M_3} \sqrt{g_{\mathrm{E}, M_3}} e^{4 A_\mathrm{E}} R_{\mathrm{E},M_3}}{\int_{M_3} \sqrt{g_{\mathrm{E}, M_3}} e^{2 A_\mathrm{E}} R_{\mathrm{E},\mathrm{AdS}_7}}\right|\gg 1\,,
\end{equation}
where $M_3$ is the internal metric, and E means Einstein frame. It might make one hopeful: on AdS$_7$ solutions with O6-planes, it can be checked that the numerator in (\ref{eq:MKK-est}) diverges, and the criterion is satisfied. Moreover, AdS$_7$ solutions with O6s seem to pass standard holographic tests \cite{apruzzi-fazzi}. However, we will now see that in fact for all the supersymmetric solutions as well as their non-supersymmetric sisters the scale of the cosmological constant is comparable
with the scale of the KK excitations. We will do so by showing that 
there are always many spin-2 modes whose mass is comparable with the AdS$_7$ cosmological constant.
Thus, the physics in AdS$_7$ is not really seven-dimensional. 

It was found in \cite{bachas-estes} that the transverse traceless
oscillations $l_{\mu \nu}$ of the AdS$_7$ metric decouple from the other
spins. Decomposing them as
\begin{equation}
  l_{\mu \nu} = \sum_i l_{\mu \nu}^i (x) \psi_i (z, \theta, \varphi) \;,
\end{equation}
where the index $i$ runs over the KK-modes. The masses of the spin-2 fields are given by the eigenvalue equation
\begin{equation}
  - \frac{e^{- 5 A + 2 \phi}}{\sqrt{g_{3}}} \partial_a \left( \sqrt{g_3}
  e^{7 A - 2 \phi} g^{a b}_3 \partial_b \psi \right) = M^2 \psi
  \label{eq:spin2Eq}, \qquad
\end{equation}
where the background has been rewritten a $ds^2_{10} = e^{2 A} ds^2_{\text{AdS}_7} + ds^2_3$, and $a,b$ are indices of $M_3$; from now on we
suppress the index $i$. We can expand the
eigenfunction $\psi$ on a basis of function on $S^2$:
\begin{equation}
  \psi \equiv \sum_{l, m} \hat{f}_{l, m} (z) Y^{l, m} (\theta, \varphi)
\end{equation}
which we take to be the eigenfunctions of the standard spherical laplacian:
\begin{equation}
  \frac{1}{\sqrt{g_{S^2}}} \partial_m \left( \sqrt{g_{S^2}} g^{m n}_{S^2}
  \partial_n Y^{l, m} (\theta, \varphi) \right) = - l (l + 1) Y^{l, m}
  (\theta, \varphi)
\end{equation}
With these definition the eigenvalue equation {\eqref{eq:spin2Eq}} becomes
\begin{equation}
  -\frac{120 X^2}{V} \frac{1}{\alpha \ddot{\alpha}}
  (X^{- 5} \partial_z (\alpha^2 \partial_z \hat{f}_{l, m}) - l (l + 1) (X^{-
  5} \dot{\alpha}^2 - 2 \alpha \ddot{\alpha}) \hat{f}_{l, m}) = - M^2
  \hat{f}_{l, m}
\end{equation}
This equation can be further simplified with the redefinition $\hat{f}_{l, m}
\equiv \alpha^l f_{l, m}$, such that the original fluctuation is expanded as
\begin{equation}
  \psi \equiv \sum_{l, m} \alpha^l (z) f_{l, m} (z) Y^{l, m} (\theta, \varphi)
\end{equation}
and we are left with the Sturm--Liouville problem
\begin{equation}
  \partial_z (\alpha^{2 l + 2} \partial_z f) = \left( M^2 \frac{V_0}{120 X^{-3}} - \frac{2 l^2 + l (2 + X^{- 5})}{X^{- 5}} \right) \alpha^{2 l
  + 1} \ddot{\alpha} f .
\end{equation}

We now immediately see that $f = \text{costant}$ is a universal solution: it is valid for any $\alpha$. This is crucial, and it will allow us to prove the absence of separation of scales for any boundary condition, brane source and flux allowed in the solutions. This leads to the masses
\begin{equation}
	M^2_\text{susy} = 8 l (2l+3) \, ,\qquad M^2_\text{non-susy} = 12 l (l+2)\,.
\end{equation}
(In fact it can be shown \cite{passias-t} that these are the lowest eigenvalues for any given $l$.) For example for $l = 0$ this gives the massless graviton, and for $l = 1$
\begin{equation}
	M^2_\text{susy} = 40 \,,\qquad  M^2_\text{non-susy} = 36 \,.
\end{equation}
We can relate these masses to mass of the cosmological constant, defined to be
to the Ricci scalar of the $\text{AdS}_7$ metric,
\begin{equation}
  M^2_{\text{AdS}_7} \equiv R_{\text{AdS}_7} = - 6 \times 7 = - 42,
\end{equation}
and we conclude that there is no separation of scales in general.

It has been conjectured \cite{lust-palti-vafa} that absence of scale separation is a generic feature of AdS solutions in quantum gravity, and the result described here confirms the conjecture for all AdS$_7$ supersymmetric solutions. 

For the non-supersymmetric case, however, we cannot claim a general proof, since we don't know how exhaustive our set of solutions is; in fact we will mention in section \ref{sub:ab} a solution with smeared D6-branes that falls outside it. 

% subsection scale (end)

\subsection{Gauged supergravity} % (fold)
\label{sub:gsugra}

In section \ref{sub:sol} we saw that every supersymmetric AdS$_7$ solution of type II admits a consistent truncation to minimal gauged supergravity in seven dimensions. This allowed us to find a non-supersymmetric sister solution. A strong limitation of this theory, however, is that all solutions are identified as the same vacuum; in a sense, minimal gauged supergravity captures a common, universal sector of all AdS$_7$ solutions. 

For various applications it is useful to be able to embed more than one vacuum in the same effective theory. In seven dimensions the only possibility is to couple gravity to vector multiplets. It is natural to choose their gauge group to be $\mathrm{SU}(k)^2$. With this choice, one obtains \cite{deluca-gnecchi-lomonaco-t} a theory with many supersymmetric vacua that appear to be in one-to-one correspondence with all type II AdS$_7$ solutions with a given $k=\sum_a a f^\mathrm{L}_a = \sum_a a f^\mathrm{R}_a$, as defined in section \ref{sub:sol}. Moreover, the theory has domain walls connecting various solutions, which nicely correspond to the RG flows one would expect from field theory \cite{gaiotto-t-6d}.

Let us first review in general how to couple the gravitational sector (\ref{eq:gra-mult}) to an arbitrary number $n_V$ of vector multiplets:
\begin{equation}\label{eq:vec-mult}
(A_\mu, \lambda^A, \phi^i)_a\,,\quad a=1,\dots, n_V\,.
\end{equation}
The scalars $\phi^i_a$ can be understood as coordinates on the coset manifold 
\begin{equation}
\mathcal{M}=\frac{\mathrm{SO}(3,n_V)}{\mathrm{SO}(3)\times \mathrm{SO}(n_V)}\,.
\end{equation}
The scalars can be thus collected in a coset representative $L^I{}_J\in \mathrm{SO}(3, N)$, where we defined the collective index $I=(i, a)$; one can then gauge a subgroup of $\mathrm{SO}(3, n_V)$. The gauged group must have dimension $3+n_V$ and structure constants $f^I{}_{JK}$ such that $f_{IJK}=\eta_{IL}f^L{}_{jk}$ are completely antisymmetric, where $\eta=\text{diag}\{-\mathbf{1}_{3\times 3}, \mathbf{1}_{n_V\times n_V}\}$. The groups satisfying such constraint are of the form $G_0\times G_{n_V}$ where $G_0$ can be $\mathrm{SO}(3)$ and $G_{n_V}$ is an $n_V$-dimensional compact group.\footnote{$G_0$ can be more in general one of the following six choices: $\mathrm{SO}(3)$, $\mathrm{SO}(3,1)$, $\mathrm{SL}(3, \mathbb{R})$, $\mathrm{SO}(2,1)$, $\mathrm{SO}(2,2)$, $\mathrm{SO}(2,1)\times \mathrm{SO}(2,2)$.} There is a potential on the scalars:
\begin{equation}
\label{eq:potgsugra}
V=\frac{1}{4}e^{-\sigma}\left(C^{iI}C_{iI}-\frac{1}{9}C^2\right)+16 h^2 e^{4\sigma}-\frac{4\sqrt{2}}{3}h e^{3\sigma/2}C\,,
\end{equation}
where now we call $\sigma$ the scalar in the gravity multiplet; $h$ is a topological mass for the two-form  $B_{\mu\nu}$  in the gravity multiplet; and we  introduced 
\begin{equation}
C=-\frac{1}{\sqrt2}f_{IJK}L^{I}{}_i L^J{}_jL^K{}_k\epsilon^{ijk}\,,\qquad C_{iI}\,=\,\frac{1}{\sqrt2}f_{IJK}L^J{}_jL^K{}_kL^R_r \epsilon^{ijk}\,.
\end{equation}
Further details about coupling minimal gauged supergravity to vector multiplets can be found in \cite{bergshoeff-koh-sezgin, bergshoeff-jong-sezgin, louis-lust,karndumri-6d,karndumri}.

As shown in \cite{deluca-gnecchi-lomonaco-t}, this theory possesses an infinite class of supersymmetric $\text{AdS}_7$ vacua of the following form:
\begin{equation}
\label{eq:svacgauged}
L^{I}{}_J\,=\,\text{exp}\left(\begin{matrix}0 & \psi\,\sigma^i_a \\ \psi\,\sigma^a_i & 0 \end{matrix}\right)\,, \quad  \tanh[\sqrt\kappa\psi]\,=\,\frac{\sqrt\kappa\,g_3}{g_{n_V}}\,,\quad e^{5\sigma/2}\,=\,\frac{g_3\,g_{n_V}}{16h\sqrt{g^2_{n_V}-g^2_3\kappa}}\,,
\end{equation}
where $g_3$ and $g_{n_V}$ are coupling constants for $\mathrm{SO}(3)$ factor and $G_{n_V}$ factor of the gauge group respectively. In \eqref{eq:svacgauged}, $\sigma^i_a T^a$ are three elements of $\mathfrak{g}_{n_V}$ forming an $\mathfrak{su}(2)$ subalgebra, with quadratic Casimir equal to $\kappa$.  

Let us now specialize to the case $n_V=2(N^2-1)$, $G_{n_V}=\mathrm{SU}(k)_L\times \mathrm{SU}(k)_R$. In this case, the embedding of $\mathfrak{su}(2)$ is defined by a pair of Young diagrams with $k$ boxes each, which it is natural to identify with $\mu^\mathrm{L,R}$ of section \ref{sub:sol}. The expectation values $\phi^i_a=\psi\sigma^i_a$ in \eqref{eq:svacgauged} can be understood as VEVs for the worldvolume scalars of the D8-branes in a given type II AdS$_7$ solution. 

Moreover, just like the minimal supergravity of section \ref{sub:sol}, each supersymmetric vacuum \eqref{eq:svacgauged} has a non-supersymmetric sister solution that only differs for the expectation value of the dilaton:
\begin{equation}\label{eq:susybr-gauged}
\frac{e^{5/2 \sigma_{\text{non-susy}}}}{e^{5/2 \sigma_{\text{susy}}}}=2\,.
\end{equation}
This is the same ratio between the non-supersymmetric and supersymmetric values of $X^5$.

While this gauged supergravity has various advantages, it is not believed to be a consistent truncation. One issue is that the action for the vectors is of the usual YM type, while the action for D-branes in ten dimensions has a DBI-type kinetic term. This manifests itself in the fact that the cosmological constants of the 7d vacua agree with those of their 10d counterpart only in the regime where the positions of the D8 are not too large with respect to $N$. Exceptional field theory also seems to suggest \cite{malek-samtleben-vallcamell} that in six and seven dimensions there are no consistent truncations involving an arbitrarily large gauge group. In spite of this shortcoming, the results of \cite{deluca-gnecchi-lomonaco-t} suggest that this theory manages to capture some of the properties of the ten-dimensional background.

% subsection gsugra (end)

% section ads (end)

\section{Perturbative instabilities} % (fold)
\label{sec:pert}

We have seen in section \ref{sub:scale} that the spectrum of spin-2 operators is given by the spectrum of a certain operator acting on internal functions. For lower spins, performing a KK reduction is made a lot harder by the presence of non-trivial warping function $A$ and dilaton $\phi$. Rather than performing this challenging computation, in this section we will compute several masses that are accessible more easily. This will be enough to reveal some instabilities. 

In section \ref{sub:specgsugra} we compute the mass matrices of the scalars in the seven-dimensional supergravity in section \ref{sub:gsugra}, extending a computation in \cite[Sec.~4.4]{deluca-gnecchi-lomonaco-t} to the non-supersymmetric case. We then reproduce the same mass matrices in section \ref{sub:nabD6} by looking at the potential for non-abelian D6-branes. In section \ref{sub:anal} we analyze the spectrum of these mass matrices and interpret it. The results indicate an instability for all solutions where some D8-branes coincide. In section \ref{sub:ab} we also look at abelian perturbations involving D8-branes.

\subsection{Scalar spectrum from gauged supergravity}% (fold) 
\label{sub:specgsugra}

The gauged supergravity reviewed in section \ref{sub:gsugra} has a set of vacua \eqref{eq:svacgauged} in one-to-one correspondence with type II AdS$_7$ solutions with given $k=\sum_a a f^\mathrm{L}_a=\sum_a a f^\mathrm{R}_a$. Moreover, it also has non-supersymmetric sister solutions, obtained by the map (\ref{eq:susybr-gauged}), which is natural to identify with the non-supersymmetric solutions (\ref{eq:nonsusy}).

In this section we will review the computation in \cite[Sec.~4.4]{deluca-gnecchi-lomonaco-t} of the masses around the supersymmetric vacua, and we will discuss how to modify it for the non-supersymmetric ones. Although this theory is not a consistent truncation, the results in \cite{deluca-gnecchi-lomonaco-t} showed it to be a good approximation if the D8 charges are not too large with respect to $N$. So the scalar masses we will compute should approximate some of the masses of the KK tower around the corresponding type II vacua. 

Recall we can divide the D8-branes in two sets: left of the massless region there are $f^\mathrm{L}_a$ D8-branes of charge $a$, and right of the massless region there are $f^\mathrm{R}_a$ D8-branes of charge $-a$. One then associates to both a Young diagram $\mu^\mathrm{L,R}$, and an $\mathfrak{su}(2)$ subalgebra $\sigma^i_{\mathrm{L,R}}$ of matrices
\begin{equation}
	[\sigma^i_\mathrm{L,R}, \sigma^j_\mathrm{L,R}]= \epsilon^{ijk} \sigma^k_\mathrm{L,R}\,
\end{equation}
which then determine the vacuum. The mass spectrum is the union of a set of L masses and a set of R masses, coming from each sector. For simplicity, in the following discussion we will then focus on one of these, say $\sigma^i\equiv \sigma^i_\mathrm{L}$. 

In general $\sigma^i$ is a reducible representation, which is associated to the rows of the Young diagram $\mu\equiv \mu^\mathrm{L}$: 
\begin{equation}\label{eq:sigma-rep}
	\sigma^i: \mathbf{d_1} \oplus  \ldots \oplus \mathbf{d_p}\,.
\end{equation}
The $d_i$ are nothing but the D6-charges and positions of the various D8-branes, which we choose to list in decreasing order. So $d_1=L$ is the D6-charge of the D8-brane with largest charge, which is located at $z=L$, the left border of the massless region; $d_2\le d_1$ is the next largest (which might coincide with $d_1$ if there are several D8s at $z=d_1=L$); and so on. D6-branes are considered in this description as D8-branes with charge 1.

In our units where $L_\mathrm{AdS_7}=1$, the Breitenlohner--Freedman (BF) bound is
\begin{equation}\label{eq:BF}
	m^2 \ge -9\,.
\end{equation}

The mass of the dilaton $\sigma$ is the easiest to compute; it is given by 
\begin{equation}\label{eq:dil-mass}
	m^2_\text{susy}= -8 \, ,\qquad m^2_\text{non-susy} = 12\,,
\end{equation}
which are both above the bound (\ref{eq:BF}).

For the other masses, following \cite{louis-lust} we can focus on fluctuations of the type
\begin{equation}
\delta L^I{}_J\,=\,\left(\begin{matrix} 0 & \delta \phi^i_a \\ \delta\phi^b_j&0 \end{matrix}\right)\left(\begin{matrix} L^j_k & L^j_c \\ L^a_k & L^a_c \end{matrix}\right)\,.
\end{equation}
Recall that the vacuum is specified by a triplets of elements $\sigma^i$ forming an $\mathfrak{su}(2)$ subalgebra. The masses of the fluctuations can be read from:
\begin{equation}
\label{eq:delta2Vs}
\delta^2 V=-8 \mathrm{Tr}\left( \delta\phi^i\delta\phi^i-2[\sigma^i, \delta\phi^j][\sigma^i, \delta\phi^j]+2[\sigma^i, \delta\phi^j][\sigma^j, \delta\phi^i]+[\sigma^i,\sigma^j][\delta\phi^i,\delta\phi^j] \right)\,
\end{equation}
around the supersymmetric vacuum, and
\begin{equation}\label{eq:delta2Vns}
	\delta^2 V = -12 \text{Tr}\left( \delta\phi^i\delta\phi^i-[\sigma^i, \delta\phi^j][\sigma^i, \delta\phi^j]+[\sigma^i, \delta\phi^j][\sigma^j, \delta\phi^i]\right)
\end{equation}
around the non-supersymmetric one.

Of course there is no guarantee that our 7d gauged supergravity really captures all the masses  one would see from a full KK reduction. For this reason, in the rest of this section we look at the problem from different points of view.

% subsection specgsugra (end)

\subsection{D6-brane polarization} % (fold)
\label{sub:nabD6}

As shown in \cite{bachas-douglas-schweigert,myers}, a stack of D$p$-branes can polarize into D$(p+2)$-branes. This can be understood using the non-Abelian action for the D$p$-stack. In this subsection we will reproduce the masses we just derived using this approach. 

Let us briefly review the action proposed in \cite{myers} to describe the dynamics of stack of $N$ Dp-branes in a given background, beyond the Abelian approximation. Let us assume the D$p$-brane to span the directions $x^\mu$, $\mu=0,\dots,p$ and let us denote the transverse directions by $y^i$, $i=1,\dots,9-p$. In the cases of interest for us, the background metric $g$ can be put in a block-diagonal form with vanishing off-diagonal entries:
\begin{equation}
g=\left( \begin{matrix}  g^\shortparallel_{\mu \nu}  & \\ & g^\bot_{ij} \end{matrix} \right)\,;
\end{equation}
moreover, the NS connection $B$ only possesses legs along the transverse directions, $B=\frac{1}{2}B_{ij}dy^i\wedge dy^j$. We will also set the worldvolume field-strength to zero. With these assumptions, the action reads:
\begin{equation}
\label{eq:nabAction}
S^{N}_{\mathrm{D}p}\,=\,-\mu_p\,\int d^{p+1}x\,\text{STr}\left (\,e^{-\phi}\sqrt{-\det g^{\shortparallel}}\sqrt{\det \Theta}-P\left[e^{\frac{1}{2\pi}\iota_Y\iota_Y}\left(e^{-B}\wedge C\right)\right]\right)\,,
\end{equation}
where $\text{STr}$ denotes the symmetrized trace and $P$ the pullback on the worldvolume. The $Y^i$ are $\mathrm{SU}(k)$ adjoint fields, or in other words anti-Hermitian matrices, which are the non-Abelian counterpart of the transverse scalars. The matrix $\Theta$ is defined as
\begin{equation}
\Theta^i{}_j\,=\,\delta^i{}_j+\frac{1}{2\pi}\left[Y^i,Y^k\right]\left( g^{\bot}_{kj}-B_{kj} \right).
\end{equation}
The non-Abelian fields also enter the Chern--Simons term trough the contraction $\iota_Y$, that can be thought of as the action of the vector $Y^i\partial_{y^i}$ on a form. For instance, given a two-form $C^{(2)}=\frac{1}{2}C_{ij} dy^i\wedge dy^j$:
\begin{equation}
\iota_Y\iota_Y C^{(2)}\,=\,Y^jY^i\,C_{ij}\,=\,\frac{1}{2}C_{ij}\left[Y^j,Y^i\right]\,.
\end{equation}
The action \eqref{eq:nabAction} is usually expanded in such a way as to keep only two-derivatives terms and at most quartic scalar interactions. We can think of the resulting action as the first-order correction to the Abelian action. 

Let us now specialize the formalism to a stack of $k$ D6-branes whose worldvolume extends along AdS$_7$ in one of the backgrounds described in \ref{sub:sol}. 
We will place a stack of $k$ D6-branes at the pole $z=0$ of an AdS$_7$ solution which has a regular point there, and study their potential. We will first look at the supersymmetric case, and then repeat the computation for the non-supersymmetric one.
                                                                                    
The transverse directions are the interval coordinate $z$ and the $S^2$ angles $(\theta, \phi)$. We can think of them as polar coordinates, and we call $y^i$ the corresponding Cartesian ones, so that for example $z^2=y_i y_i$. At the non-Abelian level, $y_i \to i Y_i$, and in particular $Z^2= - Y^2$. The parallel metric is:
\begin{equation}
g^{\shortparallel}_{\mu \nu}\,=\, e^{2A} g_{\mu \nu}^{\text{AdS}_7}\,+\,g^\bot_{ij}\,\partial_\mu Y^i \partial_\nu Y^j\,.
\end{equation}
The DBI part of \eqref{eq:nabAction} can be thus expanded as follows:
\begin{equation}
\label{eq:nabDBI}
\begin{split}
S^{N,\mathrm{DBI}}_{\mathrm{D}6}\,=\,-\mu_6\,\text{Tr}\int d^{7}x\sqrt{g^{\text{AdS}_7}}e^{7A-\phi}&\left(1+\frac{1}{2} e^{-2A} g^{\mu \nu}_{\text{AdS}_7}g^\bot_{ij}\partial_\mu Y^i \partial_\nu Y^j\right.\\
&\left.\,\,\,-\frac{1}{4\pi}B_{ij}[Y^j, Y^i]+\frac{1}{16\pi^2}[Y^i, Y^j]g^\bot_{jk}[Y^k, Y^l]g^\bot_{li}\right)\,.
\end{split}
\end{equation}
The warping, dilaton and fluxes explicitly depend on the transverse radial coordinate $z$, which has non-Abelian extension $i Z$. This also needs to be taken into account.
           
Since we assumed the solution to be regular at $z=0$, the prefactor $e^{7A-\phi}$ in \eqref{eq:nabDBI} is non-vanishing. Moreover, locally the solution has the form 
\begin{equation}\label{eq:loc-reg}
	\alpha = \frac{27}2 \pi^2 n_0 z (\zeta^2-z^2)\,.
\end{equation}
This is the same as the solution (\ref{eq:1D6}), but with the real parameter $\zeta$ now no longer having the interpretation of the length of the interval. Indeed in a solution with D8s, one still cuts a piece of (\ref{eq:loc-reg}) and glues it to some other local solution. For example we see in the solution (\ref{eq:2D8s}) with two D8 stacks that the piece of the solution for $z\in [0,\mu]$ is of the form (\ref{eq:loc-reg}) with $\zeta^2= 3 \mu (N-\mu)$. Using this in (\ref{eq:metX})--(\ref{eq:BX}) we get the approximate metric
\begin{equation}
 ds^2\,\approx\,\frac{8 L^2 \pi \zeta}{\sqrt{3X}} ds^2_{\text{AdS}_7}\,+\,\frac{2\sqrt3 \pi X^{5/2}}{\zeta}\left( dz^2+z^2 ds^2_{\text{S}^2} \right)\,,
\end{equation}
so that the internal space looks like $\mathbb{R}^3$, and
\begin{equation}
	B_{ij}\,=\,\frac{6\pi(1-6 X^5)}{3 \zeta^2}\epsilon_{ijk}Y^k\,.
\end{equation}
An analogous expression can be written for $C_9$. For the supersymmetric case we obtain the action
\begin{equation}
\label{eq:nabActionSusy}
S^{N}_{\mathrm{D6}}=-\frac{128 \pi^3}{3}n_0 \zeta^2\mu_6  \mathrm{Tr}\int d^7_x \sqrt{g_{\mathrm{AdS}_7}}\left(\frac{1}{2}\partial_\mu Y^i\partial^\mu Y^i-4 Y^i Y^i-2\epsilon_{ijk}[Y^i, Y^j]Y^k+[Y^i, Y^j]^2 \right)\,.
\end{equation}
The non-Abelian vacua of the potential are now obtained by solving
\begin{equation}
-8Y^i-6[Y^j, Y^k]\epsilon_{jk}{}^i-4[[Y^i, Y^l], Y^l]=0\,.
\end{equation}
As in \cite{myers}, we look for solutions such that $Y^i$ form an $\mathfrak{su}(2)$ algebra, $[Y^i, Y^j]\,=\, R\,\epsilon_{ijk}Y^k$, so that the vacuum equation reduces to:
\begin{equation}
-2-3R+2R^2=0
\end{equation}
admitting in particular a solution for $R=2$, so that $\frac12 Y^i$ obey the usual $\mathrm{su}(2)$ algebra. Expanding \eqref{eq:nabActionSusy} around this new vacuum $Y^i=2 \sigma^i+\delta \phi^i$:
\begin{equation}
\begin{split}
S^{N}_{\mathrm{D6}}=-\frac{512 \pi^3}{3}n_0 \zeta^2\mu_6\text{Tr}\int d^7_x \sqrt{g^{\text{AdS}_7}}&\Big(\frac{1}{2}\partial_\mu \delta \phi^i\partial^\mu \delta \phi^i-4 \delta \phi^i \delta \phi^i-4 [\sigma^i, \sigma^j][\delta \phi^i, \delta \phi^j]\\
&\,\,\,+8[\sigma^i, \delta \phi^j][\sigma^i, \delta \phi^j]-8[\sigma^i, \delta \phi^j][\sigma^j, \delta \phi^i]\Big)\,.
\end{split}
\end{equation}
The mass matrix is now given by
\begin{equation}\label{eq:mms}
\frac{-8}{\mathrm{Tr}(\delta \phi^i \delta \phi^i)}\mathrm{Tr}\left(\delta \phi^i\delta \phi^i-2[\sigma^i, \delta \phi^j][\sigma^i, \delta \phi^j]+2[\sigma^i, \delta \phi^j][\sigma^j, \delta \phi^i]+ [\sigma^i, \sigma^j][\delta \phi^i, \delta \phi^j] \right)
\end{equation}
in agreement with (\ref{eq:delta2Vs}).

In the same way, in the non-supersymmetric case the action reads:
\begin{equation}
\label{eq:nabActionNonSusy}
S^{N}_{\mathrm{D6}}=-\frac{128 \pi^3}{3}n_0 \zeta^2\mu_6\text{Tr}\int d^7_x \sqrt{g^{\text{AdS}_7}}\left(\frac{1}{2}\partial_\mu Y^i\partial^\mu Y^i-6 Y^i Y^i-\epsilon_{ijk}[Y^i, Y^j]Y^k+\frac{3}{4}[Y^i, Y^j]^2 \right)\,.
\end{equation}
Again, there exists a vacuum such that $\frac12 Y^i$ obey the $\mathfrak{su}(2)$ algebra, and expanding around it we get
\begin{equation}
\begin{split}
S^{N}_{\mathrm{D6}}=-\frac{512 \pi^3}{3}n_0 \zeta^2\mu_6\mathrm{Tr}\int d^7_x \sqrt{g_{\mathrm{AdS}_7}}&\Big(\frac{1}{2}\partial_\mu \delta \phi^i\partial^\mu \delta \phi^i-6 \delta \phi^i \delta \phi^i\\
&\,\,\,+6[\sigma^i, \delta \phi^j][\sigma^i, \delta \phi^j]-6[\sigma^i, \delta \phi^j][\sigma^j, \delta \phi^i]\Big)\,.
\end{split}
\end{equation}
so that the mass matrix is
\begin{equation}\label{eq:mmns}
\frac{-12}{\mathrm{Tr}(\delta \phi^i \delta \phi^i)}\mathrm{Tr}\left(\delta \phi^i\delta \phi^i-[\sigma^i, \delta \phi^j][\sigma^i, \delta \phi^j]+[\sigma^i, \delta \phi^j][\sigma^j, \delta \phi^i]  \right)
\end{equation}
in agreement with (\ref{eq:delta2Vns}).

So we have reproduced the mass matrices in section \ref{sub:specgsugra}. 

% subsection nabD6 (end)

\subsection{Spectrum analysis} % (fold)
\label{sub:anal}

Having computed the mass matrices (\ref{eq:mms}), (\ref{eq:mmns}) from two points of view, we now proceed to compute their spectrum and to interpret it. For the supersymmetric case this was again partially done in \cite[Sec.~4.4]{deluca-gnecchi-lomonaco-t}.

The $\delta \phi$ are valued in the adjoint of $\mathfrak{su}(N)$. We thus choose a basis $T^a$ of $\mathfrak{su}(N)$, which we take to be normalized such that $\mathrm{Tr}(T^a T^b)= \delta^{ab}$. It will transform in some reducible representation of $\mathfrak{su}(2)$:
\begin{equation}\label{eq:[sT]}
[\sigma^i, T^a]\,=\,j^i_{ab} T^b\,.
\end{equation}
From \eqref{eq:delta2Vs} and (\ref{eq:delta2Vns}), taking into account the normalization of the kinetic terms, we then have the following mass matrices:
\begin{align}
	M^{ij}_{ab,\text{susy}}&=-8\left( \delta^{ij}(1+2 j^k j^k)-2 j^{(i} j^{j)} \right)_{ab}\,,\\
	M^{ij}_{ab,\text{non-susy}}&=-12 \left( \delta^{ij}(1+ j^k j^k)- j^{j} j^{i} \right)_{ab}
	\,.
\end{align}

The $\sigma^i$ obey the $\mathfrak{su}(2)$ algebra; using the Jacobi identity, one can show that the matrices $j^{i}$ also satisfy it, $[j^i, j^j]=\epsilon^{ijk}j^k$. Our task is then reduced to determining this representation, or in other words the dimension and the number of blocks appearing in the $j^i$. The irreducible representations appearing in the $j^i$ are given by a tensor product of two copies of (\ref{eq:sigma-rep}):
\begin{equation}\label{eq:j-rep}
\begin{split}
j^i: \ (\mathbf{d_1}\oplus \mathbf{d_2}\oplus\dots \mathbf{d}_p)&\otimes (\mathbf{d}_1\oplus \mathbf{d_2}\oplus\dots \mathbf{d}_p)\,=\,\\
\,=\, &\oplus_a(\mathbf{2d_a-1}\oplus \mathbf{2d_a-3}\oplus\dots \oplus\mathbf{1})\oplus\\
&2\oplus_{a>b}(\mathbf{d_a+d_b-1}\oplus\mathbf{d_a+d_b-3}\oplus\dots \oplus\mathbf{d_a-d_b+1})
\end{split}
\end{equation}
subtracting at the end a singlet. 

For each $\mathbf{d}$ appearing in (\ref{eq:j-rep}) there is a total of $3 d$ modes. They are organized in the representations $\mathbf{d}\otimes  \mathbf{2}$, which for $d>1$ is $ \mathbf{d} \oplus \mathbf{d+2} \oplus  \mathbf{d-2} $. This happens because the vacuum defined by the $\sigma^i$ breaks both the flavor symmetry and the SU(2) of the gauge fields, but a diagonal SU(2) subgroup is preserved, which we can identify with the R-symmetry $SU(2)_\mathrm{R}$, or in other words with the angular symmetry of the $S^2$ in the 10d solution. The masses associated to these three representations is given in Table \ref{tab:masses}.

\begin{table}[h]
	\centering
\begin{tabular}{|cc|c|c|}
\hline
\multicolumn{2}{|c|}{$\mathrm{SU}(2)_{\mathcal R}$ rep.} &  $m^2_\text{susy}$ & $m^2_\text{non-susy}$\\
\hline\hline
$\mathbf{d}$   & ($d\ge 2$)& $0 $			& $0$  \\
$\mathbf{d-2}$ & ($d\ge 3$)& $4(d-1)(d+2)$ & $3(d+1)(d+5)$ \\
$\mathbf{d+2}$ & 		  & $4(d+1)(d-2)$ & $3(d+1)(d-3)$ \\
\hline
\end{tabular}
\caption{\small Masses for the supersymmetric and non-supersymmetric vacua, and their R-symmetry representation.}
\label{tab:masses}
\end{table}

As an example of the algorithm above, we can consider the vacua in \cite{karndumri-6d}, which correspond to taking $k=2$  and a single D8 at $z=2$. The vacuum is specified by $\mu=\mu^\mathrm{L}=${\tiny $\yng(1,1)$}. The $\sigma^i$ have a single block of dimension 2; so in (\ref{eq:sigma-rep}) there is only one $d=2$. According to (\ref{eq:j-rep}) and the comment below it, the $j^i$ are obtained by computing $\mathbf{2}\otimes \mathbf{2}= \mathbf{3}\oplus \mathbf{1}$; subtracting a singlet, only the $\mathbf{3}$ remains. Now we use Table \ref{tab:masses}, which associates to the $\mathbf{3}$ a $\mathbf{3}$ of $SU(2)_\mathrm{R}$ with $m^2_\text{susy}= m^2_\text{non-susy}=0$, a $\mathbf{1}$ with $m^2_\text{susy}=40$, $m^2_\text{non-susy}=36$, and a $\mathbf{5}$ with $m^2_\text{susy}=16$, $m^2_\text{non-susy}=0$. Together with the mass of the dilaton in (\ref{eq:dil-mass}), these reproduce the masses in \cite[Sec.~3.1]{karndumri-6d}.

For a more complicated example, consider now $k=3$, namely a single D8 at $z=3$, with D6-charge equal to 3. Now $\mu=\mu^\mathrm{L}=${\tiny $\yng(1,1,1)$}, and in (\ref{eq:sigma-rep}) there is only one $d=3$. Now (\ref{eq:j-rep}) says that the $j^i$ contain the two representations $\mathbf{5}$ and $\mathbf{3}$. For the non-supersymmetric case, Table \ref{tab:masses} associates to the $\mathbf{5}$: a $\mathbf{5}$ of $SU(2)_\mathrm{R}$ with $m^2_\text{non-susy}=0$, a $\mathbf{3}$ with $m^2_\text{non-susy}=96$ and a $\mathbf{7}$ with $m^2_\text{non-susy}=36$. In the same way, to the $\mathbf{3}$ we associate a $\mathbf{3}$ of $SU(2)_\mathrm{R}$ with $m^2_\text{non-susy}=0$, a $\mathbf{1}$ with $m^2_\text{non-susy}=36$, and a $\mathbf{5}$ with $m^2_\text{non-susy}=0$.

We now try to interpret the results, starting from the supersymmetric case. All masses in Table \ref{tab:masses} are above the bound (\ref{eq:BF}), as expected. The first row, $m^2=0$, was discussed at length in \cite[Sec.~4.4]{deluca-gnecchi-lomonaco-t}: via the usual correspondence
\begin{equation}\label{eq:Dm}
	\Delta (\Delta-6)= m^2
\end{equation}
they correspond to marginal operators ($\Delta=6$). They however do not represent genuine massless deformations, which are forbidden for six-dimensional SCFTs, but rather NG bosons. 

The other two entries are more interesting. Via (\ref{eq:Dm}) they correspond to operators of dimensions
\begin{equation}\label{eq:D-susy}
	\Delta=2d+4 \, ,\qquad \Delta=2d+2
\end{equation}
respectively. They have the correct dimensions and $\mathrm{SU}(2)_\mathrm{R}$ representation for the scalars in a vector multiplet representation of $\mathfrak{osp}(6,2|1)$, the superconformal algebra for ${\cal N}=(1,0)$ 6d SCFTs and AdS$_7$ solutions; see for example \cite[Table 1]{passias-t}, where the scalars are recognized as having weights $(\Delta;0,0,0)$; $p$ there is to be identified with $d+1$ here. The case $d=1$ in our Table \ref{tab:masses} is peculiar because only the last row of Table \ref{tab:masses} survives, giving $m^2_\text{susy}=-8$ and $\Delta=4$. This corresponds to the case $p=2$ in \cite[Table 1]{passias-t}, which is the massless vector multiplet. Indeed if we count the number of such $d=1$ scalars by using (\ref{eq:j-rep}) we get 
\begin{equation}\label{eq:dim-flavor}
	-1 +\sum_a f_a^2 \,,
\end{equation}
which is the dimension of the residual gauge symmetry
\begin{equation}\label{eq:flavor}
	\mathrm{S}\left(\Pi_a U(f_a)\right)\,.
\end{equation}
In the dual SCFT$_6$, these $d=1$, $m^2_\text{susy}=-8$, $\Delta=4$ operators are the lowest component in a flavor supercurrent.

Recall also that scalars in six dimensions have dimension 2; this already manifested itself in the spin-2 analysis \cite{passias-t} reviewed in section \ref{sub:scale}. The dimensions in (\ref{eq:D-susy}) can then be explained in terms of mesons in the dual quiver description of \cite{hanany-zaffaroni}. As detailed in \cite{gaiotto-t-6d,cremonesi-t}, D8-branes are dual to flavors in the quiver; a meson in the bifundamental of $\mathrm{SU}(f_a) \times \mathrm{SU}(f_b)$ has a minimum number of scalars of order $a-b$, which corresponds to the $d_a-d_b+1$ in the second line of (\ref{eq:j-rep}). The $d_a+d_b+1$ corresponds then to a chain of mesons that starts from flavor $f_a$ goes to the end of the quiver and comes back to flavor $f_b$. The dimensions on the first line have a similar interpretation as mesons in the adjoint of a single $\mathrm{SU}(f_a)$. Coming back to the gravity solution, these mesons can be understood as strings connecting the various D8-branes. 

We now turn to the non-supersymmetric case. In this case, there is no expectation that the masses and the dimensions of the dual operators should arrange themselves according to dimensions of a superconformal group. Indeed the KK towers of Table \ref{tab:masses} give rise to non-rational numbers that do not lend themselves to an immediate interpretation. They are likely to still be interpreted as mesons, but now there is no BPS protection against binding energy, and so the dimensions are not expected to be simply the sum of the dimensions of the constituent scalar fields.

So we get back to our main purpose, which is stability. For non-supersymmetric vacua, we see that the only negative masses come from the last row in the Table: 
\begin{equation}\label{eq:below-BF}
	d=1 \to m^2_\text{non-susy}= -12 \, ;\qquad d=2 \to m^2_\text{non-susy}=-9.
\end{equation}
Since $-12$ is below the bound (\ref{eq:BF}), we see an instability. However, this value is not present for all solutions: the singlet ($d=1$) might simply be absent, as in the example $\mu^\mathrm{L}=${\tiny $\yng(1,1,1,1)$} above. To see when this happens, we can notice that the $m^2=-12$ modes are those with $d=1$ in the last line of Table \ref{tab:masses}, which had $m^2=-8$ and $\Delta=4$ in the supersymmetric case. They were associated with the unbroken symmetry (\ref{eq:flavor}), with dimension (\ref{eq:dim-flavor}). So we conclude that this instability is absent only for solutions that have a single D8, for which the (\ref{eq:flavor}) would be $\mathrm{S}(\mathrm{U}(1))$.

To see this instability more clearly we can consider an example with two coincident D8-branes of charge $d$, which corresponds to a $\sigma^i$ containing two identical blocks of size $d$; so we can write $\sigma^i= \sigma^i_a \otimes 1_2$, where the subscripts denote the size of the matrices. This is the representation $\mathbf{d} \oplus \mathbf{d}$. If we deform this by $\delta \phi^i = 1_a \otimes \sigma^i_2$, we are separating the two D8s. Indeed $\sigma^i + \epsilon \delta \phi^i$ becomes an $\mathfrak{su}(2)$ representation for $\epsilon=1$; looking at its $i=3$ component we see that it corresponds to the representation $\mathbf{d+1} \oplus \mathbf{d-1}$. Moreover, since $[\sigma^i, \delta \phi^j]=[\sigma^i_a \otimes 1_2, 1_a \otimes \sigma^j_2]=0$, we see from (\ref{eq:[sT]}) that the $j^i=0$; so this is a $d=0$ component of (\ref{eq:j-rep}). Table \ref{tab:masses} tells us that these modes are a triplet of $\mathrm{SU}(2)_\mathrm{R}$ with $m^2_\text{susy}=-8$, and $m^2_\text{non-susy}=-12$. So in this case we see an instability that separates the D8s. More generally, whenever there are two blocks one is able to find another triplet $\delta \phi^i$ that commutes with the $\sigma^i$, which by (\ref{eq:[sT]}) gives $j^i=0$ and then $d=0$ in (\ref{eq:j-rep}). 

Recall that there are in fact two sets of masses, one coming from the L sector (left of the massless region), and one coming from the R sector (right of the massless region). In the discussion so far we have focussed on the L sector, but the same conclusion holds for the R sector as well. So there should be a single D8 on both sides. This leaves us with the solution (\ref{eq:2D8s}) with $n=1$.

There is actually also the possibility that no massless region exists at all. This regime is far from the small-D8 assumptions we have made in both sections \ref{sub:specgsugra} and \ref{sub:nabD6}. However, the mass $-12$ is well below the bound of $-9$ in (\ref{eq:BF}), and even if it receives corrections far from our regime, it is reasonable to expect this instability to hold in general.  

We thus conclude
\begin{equation}\label{eq:D8-inst}
	m^2 \supset -12  \ \Leftrightarrow \ \# \mathrm{D8_L} \,>1\,, \# \mathrm{D8_L} \,>1\,\,.
\end{equation}

Notice that this result is consistent with the value of the cosmological constant for these vacua. In the Young diagram language we briefly reviewed in section \ref{sub:sol}, the stable solutions are those where $\mu^\mathrm{L}$ and $\mu^\mathrm{R}$ are both horizontal. 
In the supersymmetric case, the dual SCFTs are at the bottom of a hierarchy of RG flows, since horizontal Young diagrams are the largest in the hierarchy of nilpotent elements. One then expects their $a$ Weyl anomaly to be the largest at a fixed value of $k$, and the cosmological constants of the corresponding vacua to be smallest. In the non-supersymmetric case, it makes sense that the instabilities are driving the system to the vacuum with the smallest cosmological constant. 

In the next section we will show that the solution (\ref{eq:2D8s}) with $n=1$ suffers from a non-perturbative instability. However, it may also have a perturbative instability due to long strings stretching between the left and the right D8-branes. This would be consistent with the instability being associated to scalars of unbroken flavor currents. For instance, the flavor symmetry of the dual theory is actually $\mathrm{S}(\mathrm{U}(1)_\mathrm{L} \times \mathrm{U}(1)_\mathrm{R})$ instead of $\mathrm{SU}(1)_L \times \mathrm{SU}(1)_R$. On the other hand, these instabilities would potentially be visible in a full perturbative string theory analysis. From our prospective, the 7d supergravity coupled to vector multiplets does not contain the modes associated to these strings connecting the left with the right D8-branes and, in particular, it is not valid in the regime where the left vectors mix with the right ones. Therefore we do not see the possible instability associated to these long strings. 

Finally, we also expect that a similar perturbative analysis holds for the solutions with O6$^{\mp}$ planes. In that case, the Higgs branch RG-flows are still classified by a nilpotent orbits of $\mathfrak{so}$ or $\mathfrak{usp}$ Lie-algebras, which are again specified by Young diagrams. We also notice that the formula of the masses \eqref{eq:j-rep} does not explicitly depend on the details of the 7d supergravity gauge group, $G_{n_V}$. According to the interpretation of the modes with masses $m_{\rm susy}=-8$ and $m_{\rm non-susy}=-12$ as scalar in an unbroken flavor current of the dual theory (for the supersymmetric case), it seems natural to expect that when the supersymmetric solution is dual to a theory with nontrivial flavor symmetry, its non-supersymmetric sister is perturbatively unstable. Solutions with O8 planes are a bit different. The dual RG-flows in the Higgs branch are parametrized by discrete holonomies for the flavor symmetry \cite{mekareeya-ohmori-tachikawa-zafrir,frey-rudelius}, instead of continuous $\mathfrak{su}(2)$ nilpotent orbits. It would be interesting to generalize our setup to include these cases as well. We plan to come back to this in the future.

% subsection anal (end)

\subsection{Abelian perturbations} % (fold)
\label{sub:ab}

The results of the previous subsections have shown that most solutions have a perturbative instability. In particular in section \ref{sub:nabD6} we have reproduced the gauged supergravity computations by considering perturbations around non-Abelian D6-brane vacua. In this section we will consider abelian brane fluctuations. 

D6-brane perturbations were already considered in \cite{danielsson-dibitetto-vargas-swamp}, and we only review them here. The action is the same as in (\ref{eq:nabAction}) for $p=6$, but now we take $N=1$ and assume the background solution already has some D6-branes at $z=0$. The non-Abelian features disappear and we are left with
\begin{equation}
	S_{\mathrm{D6}} = -\mu_6 \int d^7 x \sqrt{g_\mathrm{AdS_7}}\left(e^{7A-\phi}\sqrt{1+e^{2Q-2A}\partial_\mu z\partial^\mu z}-C_7\right)
\end{equation}
where we now omit the pullback symbol and the partial derivatives are taken with respect to the $\mathrm{AdS}_7$. The potential satisfies 
\begin{equation}
	 dC_7=F_8=-* F_2 \equiv f_8\,dz\wedge\mathrm{vol}_{\mathrm{AdS}_7}  \,, 
\end{equation}
where $F_2= f_2 \mathrm{vol}_{S^2}$ can be read off from (\ref{eq:BX}). Assuming the transverse fluctuations to be small, one can expand the action in order to have a two-derivative action:
\begin{equation}
S_{\mathrm{D6}}=-\mu_6 \int d^{7}x \,\frac{1}{2}K_{\mathrm{D6}}\left(\partial_\mu z\partial^\mu z\,+\,V_{D6}\right)\,,
\end{equation}
where we defined
\begin{equation}
K_{\mathrm{D6}}= e^{5A-\phi+2Q}|_{z_0}\,,\quad V_{\mathrm{D6}}= -C_7+e^{7A-\phi}\,.
\end{equation}
Rather than looking for a universal expression for the potential in terms of $\alpha$, we can simply compute the force $K_{\mathrm{D6}}F_z=-\partial_z(e^{7A-\phi}) + f_8$. With the boundary conditions $\alpha=0$, $\ddot \alpha\neq0$, which indicate presence of D6s, this $F_z$ is zero at $z=0$ (even before substituting the values $X=1$ and $2^{-1/5}$ from (\ref{eq:X-vac})), indicating that the probe D6 is at rest if it sits together with the pre-existing D6s. The mass can now be computed as $-\partial_z F_z|_{z=0} = \frac{120}{X^{-3} V_0}$, which from (\ref{eq:Vgs}), (\ref{eq:X-vac}) gives
\begin{equation}\label{eq:D6-inst}
	m^2_\text{susy}=-8 \, ,\qquad m^2_\text{non-susy}= -12\,. 
\end{equation}
Once again, in the non-supersymmetric case this indicates an instability, sitting below the BF bound (\ref{eq:BF}) of $-9$.

For a vacuum with several D6-branes, from the previous subsections we already know of another instability with the same mass: since D6-branes are indistinguishable from D8-branes with D6-charge equal to one, we can apply the criterion (\ref{eq:D8-inst}). These are two different instabilities, presumably leading to different fates. The instability (\ref{eq:D8-inst}) leads $k$ D6-branes to polarize into a single D8 with a D6-charge equal to $k$. On the other hand, the instability (\ref{eq:D6-inst}) has to do with abelian motions of the D6-branes; it is natural to think that it leads to a solution where the D6-branes are spread all over the internal $M_3$. 

In fact one such solution was already found (see for example  \cite[Sec.~2.2]{blaback-danielsson-junghans-vanriet-wrase-zagermann-2}), by approximating the D6 distribution as smeared.\footnote{The smearing of the sources breaks supersymmetry, as found in \cite{danielsson-dibitetto-fazzi-vanriet}.} This means that
\begin{equation}\label{eq:bianchiF2sm}
	dF_2 - H F_0 = 2\pi\delta_6\,,
\end{equation}
where $\delta_6$ is a density of D6-branes, integrating to $k$ over $M_3$. The fields are given by
\begin{subequations}\label{eq:smD6}
	\begin{align}
		d  s^2_{10} = R^2 (12 d  s^2_{\text{AdS}_7} + d 
	  s^2_{S^3}), \qquad e^{\phi} \equiv g_s = \frac{2}{R \sqrt{7}} \frac{1}{| F_0
	  |} \;,\\
	H \equiv h_1 \text{vol}_{S^3} = -R^2 \frac{5}{\sqrt{7}} \text{sign} (F_0)
	  \text{vol}_{S^3}, \qquad F_2 = 0.
	\end{align}
\end{subequations}
The only free parameters are $F_0$ and $R$, which are both quantized, the latter by imposing flux quantization $N\equiv -\frac1{4\pi^2}\int_{M_3}H=\frac5{2\sqrt7}\mathrm{sign}(F_0) R^2$. 
By integrating (\ref{eq:bianchiF2sm}) we also see that $k=n_0 N$. 

In \cite{blaback-danielsson-junghans-vanriet-wrase-zagermann-2} it was found that the solution (\ref{eq:smD6}) is perturbatively stable, albeit in a setup where only a very limited number of modes was considered. We are not aware of any other reasons it might have tachyons. In section \ref{sub:bub-sm} we will analyze its non-perturbative stability.

We now consider abelian perturbations of the D8s. We consider a probe D8 with D6-charge $p$ wrapping the $S^2$. The action is the same as in (\ref{eq:nabAction}) for $p=8$ and $N=1$; moreover, we have to perform the shift $B_2\,\rightarrow\,\mathcal{F}=B_2+2\pi f_2$:
\begin{equation}
		S_{\mathrm{D8}} = -\mu_8 \int d^9 x \sqrt{g_{S^2}}\sqrt{g_\mathrm{AdS_7}}\left(e^{7A-\phi}\sqrt{1+g_{zz}e^{-2A}\partial_\mu z\partial^\mu z}\sqrt{\frak{f}^2+e^{4Q}}-C_9+\mathcal{F}\wedge C_7 \right)\,,
\end{equation}
where $B_2=b_2 \mathrm{vol}_{S^2}$ and $\mathcal{F}=\frak{f}\,\mathrm{vol}_{S^2}= (b_2+\pi p) \mathrm{vol}_{S^2}$. For simplicity let us restrict to the case where $z$ fluctuates only along $\mathrm{AdS_7}$. The potential now satisfies 
\begin{equation}
	dC_9-H\wedge C_7=F_{10}\equiv f_{10}\,dz \wedge\mathrm{vol}_{\mathrm{AdS}_7}\wedge \mathrm{vol}_{S^2}\,.
\end{equation}
As we did before, we can assume the fluctuations of the transverse coordinate to be small in such a way to have an ordinary two-derivative action:
\begin{equation}
S_{\mathrm{D8}}=-\mu_8 \int d^{7}x \left(\frac{1}{2}K_{\mathrm{D8}}\,\partial_\mu z\partial^\mu z\,+\,V_{D8}\right)
\end{equation}
with:
\begin{equation}
K_{\mathrm{D8}}= e^{5A-\phi+2Q}\sqrt{\mathfrak{f}+e^{4P}}|_{z_0}\,,\quad V_{\mathrm{D8}}=-C_9+\mathfrak{f}\,C_7+e^{7A-\phi}\sqrt{\mathfrak{f}+e^{4P}}\,.
\end{equation}
The force acting on the probe can be computed as $K_{\mathrm{D8}}\,F_z=-\partial_z(e^{7A-\phi}\sqrt{\frak{f}^2+e^{4P}})+f_{10}-\frak{f}f_8$. For any $\alpha$,  the force has three zeros: two at the boundaries of the interval and one in correspondence to the only stable point $z=p$. We can compute the mass of the fluctuations around this point evaluating $-K_{\mathrm{D8}}\,\partial_z F_z|_{z=p}\,=\,-\frac{120(1+4X^5)}{X^{-5}V_0}$. We get
\begin{equation}
m^2_\text{susy}=40 \, ,\qquad m^2_\text{non-susy}= 32\,.
\end{equation}
Notice that this value is universal, in that it does not depend on the D6-charge $p$ nor on the choice of $\alpha$ (and hence of AdS$_7$ solution).

One can similarly evaluate the fluctuation spectrum when $z$ is taken to be also a function of the $S^2$. The modes organize themselves in spherical harmonics $Y_{lm}(\theta,\phi)$, and the masses are again universal:
\begin{equation}
	m^2_\text{susy} = 40 + 16 l(l+1) \, ,\qquad m^2_\text{non-susy} = 32 + 12 l (l+1)\,.
\end{equation}

% subsection ab (end)

% section pert (end)

\section{Bubble creation} % (fold)
\label{sec:bubbles}

In the previous section we have concluded that most solutions have a perturbative instability; the one type of solution that remains is the one with two D8s, corresponding to (\ref{eq:2D8s}) with $n=1$. 

We will now discuss non-perturbative instabilities. We begin in section \ref{sub:bubbles} with a review of how this works in general, comparing various approaches, notably those in \cite{maldacena-michelson-strominger,coleman-deluccia,brown-teitelboim}.\footnote{We have benefited from discussions with K.~Eckerle, in particular about the shape of the bubbles obtained in \cite{maldacena-michelson-strominger}.} Then in section \ref{sub:ns5} we apply the formalism and show that a quantum effect can create an NS5 bubble, which then expands and makes the solutions with two D8s decay.

\subsection{Review of non-perturbative instabilities} % (fold)
\label{sub:bubbles}

Consider first quantum mechanics, and a particle in a false vacuum $x_\mathrm{f}$ of a potential $V$. The particle can tunnel through a potential barrier and end up in a true vacuum $x_\mathrm{t}$. The probability for this decay to happen can be computed using the path integral formalism \cite{coleman1,callan-coleman}; the leading contribution comes from the classical action of a trajectory of the particle in Euclidean time $\tau$. The Euclidean equation of motion is the same as the one describing the particle's motion in the inverted potential $-V$. The trajectory consists in the particle leaving the false vacuum $x_\mathrm{f}$, going through the maximum of $V$ (minimum of $-V$), arriving at some position $x_\mathrm{tunnel}$, and going back to $x_\mathrm{f}$. This is called a ``bounce''. For reviews see for example \cite[Chap.~7.2.4]{coleman} or \cite[Chap.~1.4]{marino}. 
An example of such a bounce is given by the trajectory in Fig.~\ref{fig:ns5pot}, which will be relevant for us later.

In quantum field theory, this tunnel effect cannot happen everywhere at once, as the corresponding instanton would have infinite action. Rather, what can happen is that a region of true vacuum can be suddenly created \cite{coleman1,callan-coleman}. To describe this, one again continues to Euclidean time, but the instanton now is a solution of the Euclidean field equations, still called a bounce, where a ball of true vacuum in Euclidean spacetime is surrounded by a sea of false vacuum. It can be argued that the solution with lowest action, which thus gives the largest contribution $e^{-S}$ to the path integral, is spherically symmetric. To describe this bounce one often resorts to an approximation where the fields change suddenly at the surface of a ``thin wall'' separating false and true vacua.
The presence of gravity impacts these computations, sometimes making an AdS false vacuum stable \cite{coleman-deluccia}; this is what protects supersymmetric AdS vacua from decay \cite{gibbons-hull-warner,cvetic-griffies-rey}. For a scalar field theory with a potential, this happens when the barrier between the two vacua is much larger than the vacuum energy difference.\footnote{The condition is not so clear away from the thin-wall approximation, but \cite{amsel-hertog-hollands-marolf,danielsson-dibitetto-vargas-dw} uses fake superpotentials to exclude the presence of bubbles.} 

In string theory, the bounce can be a brane. This can be viewed in two ways, roughly corresponding to the quantum mechanical and QFT points of view mentioned in the previous two paragraphs. The first point of view is as a process in Euclidean time where a brane gets created, expands, reaches a maximum size, and recollapses. This was used in \cite{maldacena-michelson-strominger}. The second point of view, more similar to Coleman--De Luccia (CDL) \cite{coleman-deluccia}, is to think of the brane as a  wall, a naturally thin one, which sources a change in a flux quantum, which in turn causes the vacuum to change as one crosses it. We will now compare the two approaches, before applying them to our set of vacua.

\subsubsection{Quantum mechanics approach} % (fold)
\label{ssub:mms}

Euclidean AdS$_d$ can be described as the hyperboloid 
\begin{equation}\label{eq:hyp}
	 -X_0^2+\sum_{i=1}^{d-1}X_i^2 + X_{d}^2=-1\,
\end{equation}
with the metric
\begin{equation}\label{eq:ds-hyp}
		 ds^2_\mathrm{EAdS}=-dX_0^2+\sum_{i=1}^{d-1}dX_i^2 + dX_{d}^2\,.
\end{equation}
In \cite{maldacena-michelson-strominger} this is parametrized as
\begin{equation}\label{eq:Xi}
	X_0 = \cosh \rho \cosh \tau \, ,\qquad 
	X_{d} = \cosh \rho \sinh \tau \, ,\qquad X_i = \sinh \rho \omega_i\,,
\end{equation}
where $\sum_i^{d-1}\omega_i^2=1$. This leads to
\begin{equation}\label{eq:glo-eads}
	ds^2_\mathrm{EAdS}= \cosh^2 \rho d\tau^2 +  d\rho^2 + \sinh^2 \rho ds^2_{S^{d-2}}\,.
\end{equation}
Consider now a brane with action
\begin{equation}\label{eq:S-brane}
	S= - T \int d^d \sigma \sqrt{g} + Q \int C_{d-1}\,;
\end{equation}
$T$ represents a tension, while $Q$ is a charge under a $(d-1)$-potential $C_{d-1}$. Its flux $F_d=d C_{d-1}$ is proportional to the volume form of EAdS: 
\begin{equation}
	F_d= f \mathrm{vol}_\mathrm{EAdS}= f \cosh \rho \sinh^{d-2} \rho  d\tau \wedge d \rho \wedge \mathrm{vol}_{S^{d-2}}\,
\end{equation}
with $f$ a constant. The potential can then be taken to be $C_{d-1}= -\frac1{d-1}f \sinh^{d-1} \rho d \tau \wedge \mathrm{vol}_{S^{d-2}}$. Let now the brane wrap $S^{d-2}$. The action is then\footnote{In evaluating the pullback of $C_{d-1}$, there is a subtle sign due to the orientation of the cycle we are integrating on.}
\begin{equation}\label{eq:S-Ebrane}
	S= -\mathrm{Vol}_{S^{d-2}} \int d\tau \left( T\sinh^{d-2} \rho\sqrt{\dot\rho^2 + \cosh^2 \rho} - \frac Q{d-1} f \sinh^{d-1} \rho\right) \,,
\end{equation} 
where $\mathrm{Vol}_{S^{d-2}}$ is now the volume of $S^{d-2}$. (We have ignored the internal dimensions here; for example this brane could be fully wrapped on the internal space, or point-like. In the latter case one would have to take into account the motion in the internal dimensions; we will comment on these subtleties when they become important in the next section.)  

(\ref{eq:S-Ebrane}) is a quantum-mechanical action for the single variable $\rho=\rho(\tau)$, already Wick-rotated to Euclidean signature, and can be studied with the methods outlined at the beginning of this section. Analytic solutions were found in \cite{maldacena-michelson-strominger} using conservation of Euclidean energy. A finite-action solution only exists if
\begin{equation}\label{eq:q}
	q \equiv \frac{Q f}{T(d-1)}>1\,,
\end{equation}
and it reads
\begin{equation}\label{eq:bounce-mms}
	\cosh \tau \cosh \rho = \cosh \rho_\mathrm{max}= \frac{q}{\sqrt{q^2-1}} \,.
\end{equation}
As the name implies, $\rho_\mathrm{max}$ is the maximum value of the radius of the bubble, attained at $\tau=0$. So \cite{maldacena-michelson-strominger} conclude that for $q>1$ a bubble can get created, destabilizing the vacuum. For $q=1$ the solution is $e^{\tau}=\cosh \rho$; for $q<1$, $\sinh \tau \sinh \rho = \sinh \rho_\mathrm{max}$. Neither of these last two has a finite action. So we see that for $q\le 1$ a bubble can not be nucleated; as promised, gravity can sometimes stabilize a false vacuum.

If $q>1$, there is a finite probability of a bubble being nucleated. This is a sphere $S^{d-2}$ at a fixed radius $\rho_\mathrm{max}$ in the global coordinates of AdS:
\begin{equation}\label{eq:glo-ads}
	ds^2_\mathrm{AdS}= -\cosh^2 \rho dt^2 +  d\rho^2 + \sinh^2 \rho ds^2_{S^{d-2}}\,,
\end{equation}
which appears at a given time. The subsequent evolution can be modelled with the classical equations of motion. Since (\ref{eq:glo-ads}) is the $t= i\tau$ Wick-rotation of (\ref{eq:glo-eads}), it makes sense to expect that the trajectory is obtained from (\ref{eq:bounce-mms}) with the same rotation; this leads to $\cos t \cosh \rho = \cosh \rho_\mathrm{max}$. We see that the bubble reaches $\rho=\infty$ at $t=\pi/2$.

% subsubsection mms (end)

\subsubsection{Manifest $\mathrm{SO}(d)$} % (fold)
\label{ssub:hyb}

We now want to interpret the computation we just saw in the spirit of CDL \cite{coleman-deluccia}, namely as a thin wall separating two vacua. A first obstacle would appear to be that CDL argue such a wall to be spherically symmetric, while (\ref{eq:bounce-mms}) doesn't seem to be. However, that's a coordinate artifact: indeed we notice that it can be rewritten in terms of the embedding coordinates (\ref{eq:Xi}) as 
\begin{equation}
	X_0 = \cosh \rho_\mathrm{max}\,.
\end{equation}
Looking at (\ref{eq:hyp}), we see that this locus is a sphere, left invariant by a $\mathrm{SO}(d)$ subgroup of the $\mathrm{SO}(d,1)$ isometry group of EAdS. (The solutions with $q=1$ and $q<1$ can similarly be written as $X_{d}=\sinh \rho_\mathrm{max}$ and $X_0=X_{d}$, corresponding respectively to a hyperboloid and paraboloid.)

To make this spherical symmetry even more manifest, we can choose a different set of coordinates:
\begin{equation}\label{eq:Xi-alt}
	X_0 = \cosh r \, ,\qquad X_I = \sinh r \Omega_I\,,
\end{equation}
where now $I=1,\,\ldots,\,d$ and $\sum_{I=1}^d \Omega_I^2=1$ parameterize an $S^{d-1}$. Now (\ref{eq:ds-hyp}) gives
\begin{equation}\label{eq:AdS-r}
	ds^2_\mathrm{EAdS}= dr^2 + \sinh^2 r ds^2_{S^{d-1}}\,.
\end{equation}
Assuming $\mathrm{SO}(d)$ symmetry from the beginning as in \cite{coleman-deluccia}, and using these new coordinates, the computation is now faster. The flux is $F_d= f \mathrm{vol}_\mathrm{EAdS}= \sinh^{d-1} r dr \wedge \mathrm{vol}_{S^{d-1}}$, and we take the potential $C_{d-1} = f c(r) \mathrm{vol}_{S^{d-1}}$, where $c'(r)= \sinh^{d-1}r$.
We take the brane to be located at some fixed value of $r$; the action now is simply
\begin{equation}\label{eq:hyb}
	S = \mathrm{Vol}_{S^{d-1}} \left(- T \sinh^{d-1} r + Q f c(r) \right)\,.
\end{equation}
Solving the equations of motion now simply means extremizing it, $\frac{\partial S}{\partial r}=0$: this gives $0=(-(d-1) T \cosh r \sinh^{d-2} r + Q f\sinh^{d-1}r)_{r=r_0}$, or in other words
\begin{equation}\label{eq:bounce-hyb}
	\tanh r_0 = \frac{(d-1)T}{Q f}= \frac 1q\,.
\end{equation}
Since from (\ref{eq:Xi-alt}) and (\ref{eq:Xi}) we have $\cosh r = \cosh \rho \cosh \tau$, this reproduces the radius $\rho_\mathrm{max}$ in (\ref{eq:bounce-mms}). Moreover, we see again that a solution only exists if $q>1$.

The computation leading to (\ref{eq:bounce-hyb}) is still not quite the same as in \cite{coleman-deluccia}. We have taken from them the assumption of spherical symmetry, but they consider walls obtained as field configuration in a scalar field theory. In their thin wall approximation, their action would be of the form
\begin{equation}\label{eq:cdl}
	\Delta S = - \mathrm{Vol}_{S^{d-1}} T \sinh^{d-1} r + \mathrm{Vol}_{B_d(r)}\Delta V\,. 
\end{equation}
We write $\Delta S$ because this is in fact the difference between the action of the false vacuum and the action of the bounce. The first term of (\ref{eq:cdl}) is as in (\ref{eq:hyb}): the tension $T$ of the wall comes from the integral over the wall of the kinetic energy. The second term in (\ref{eq:cdl}) is the volume of a ball $B_d(r)$ (with boundary $S^{d-1}$), multiplied by the difference of the energy densities in the true and false vacuum.

Our ``hybrid'' approach (\ref{eq:hyb}) is more convenient for brane bubbles in string theory, but the CDL point of view (\ref{eq:cdl}) also applies, in the following sense. \cite{coleman-deluccia} worked out the decay for scalar field theories. Superficially branes   are not obtained as domain walls in a scalar potential, but in fact that point of view does apply to string field theory; for example a D8 can be obtained as a domain wall in a tachyon for an unstable non-BPS D9 \cite{sen-nonbps}. Perhaps more concretely, the CDL point of view was applied  in \cite{brown-teitelboim} to Lagrangians where the scalar is replaced by a $d$-form field-strength, just like the one we have been using; a similar formalism, less dependent on a particular Lagrangian, was applied in \cite{harlow-ads}, focusing on AdS to AdS transitions. The two actions (\ref{eq:hyb}) and (\ref{eq:cdl}) can now be related by rewriting
\begin{equation}\label{eq:diff}
	Q\mathrm{Vol}_{S^{d-1}} f c(r)= \int_{S^{d-1}}C_{d-1}= \int_{B_d(r)} F_d\,.
\end{equation}
The difference in vacuum energies can now be related to the variation of $F$ by using the equations of motion. 

Finally let us comment on what happens to the bubble after is nucleated. Again one should now solve the Lorentzian-signature equations of motion. A simple solution is provided by an analytic continuation of the bounce itself; since the bounce was a sphere, preserving an $\mathrm{SO}(d)$, the Lorentzian solution is a hyperboloid, preserving an $\mathrm{SO}(d-1,1)$.

% subsubsection hyb (end)

\medskip

In all approaches, the relevant criterion is $q>1$ for (\ref{eq:hyb}). This is sufficient for bubble nucleation, and for the bubble to expand after being created.

It was pointed out in \cite{ooguri-vafa-ads} that the existence of a brane with $q>1$ is a natural extension to branes of the weak gravity conjecture \cite{arkanihamed-motl-nicolis-vafa}. This suggests that a decay channel of this type always exists. In the next section, we will check this expectation for our AdS$_7$ solutions. 

% subsection bubbles (end)

\subsection{NS5 bubbles for the solutions with D8s} % (fold)
\label{sub:ns5}

Our bubble will be an NS5 which extends along six directions in AdS, and which is pointlike in the internal $M_3$. We will analyze whether it can nucleate, and destabilize the vacuum. While this brane does not divide spacetime into an ``inside'' and ``outside'', it still interpolates between two AdS$_7$ solutions, that differ by one unit of the flux quantum $N$.\footnote{NS5-brane bubbles were also considered in \cite{kachru-pearson-verlinde,gautason-truijen-vanriet,danielsson-gautason-vanriet}; but in those cases the NS5 arises as a possible polarization of a D-brane. For example in \cite{kachru-pearson-verlinde} is argued to polarize into an NS5, which subsequently undergoes a tunnel effect to a pole of the internal sphere; this is described by an NS5 bubble. Here the NS5 is not present in the solutions, and its only role is to change the internal NSNS flux.}

As we commented below (\ref{eq:S-Ebrane}), the fact that the NS5 is pointlike in the internal space gives rise to an additional complication. In the approach of \cite{maldacena-michelson-strominger}, the motion is not only along $\rho$ but also along the internal directions; for simplicity we consider only the motion along $z$. 

Recall that the solutions whose stability we are investigating have two D8s, and a massless region between the two. We will assume the NS5 is in this massless region. This simplifies the analysis: in the massive regions, there would also be D6s ending on the NS5 and on one of the D8s.

So the relevant flux, which we called $F_d$ in section \ref{sub:bubbles}, is 
\begin{equation}
	H_7 \equiv e^{-2 \phi} * H_3  = h_7  \mathrm{vol}_\mathrm{AdS_7} \, ,\qquad h_7 \equiv b_2'(z) \frac{e^{7A-2 \phi}}{g_{zz}^{1/2} g_{\theta \theta}} \,;
\end{equation}
its potential is
\begin{equation}
	B_6=-\frac16 h_7 \sinh^6\rho d\tau\wedge \mathrm{vol}_{S^5} \, .
\end{equation}

\begin{figure}[ht]
	\centering
		\includegraphics[height=3in]{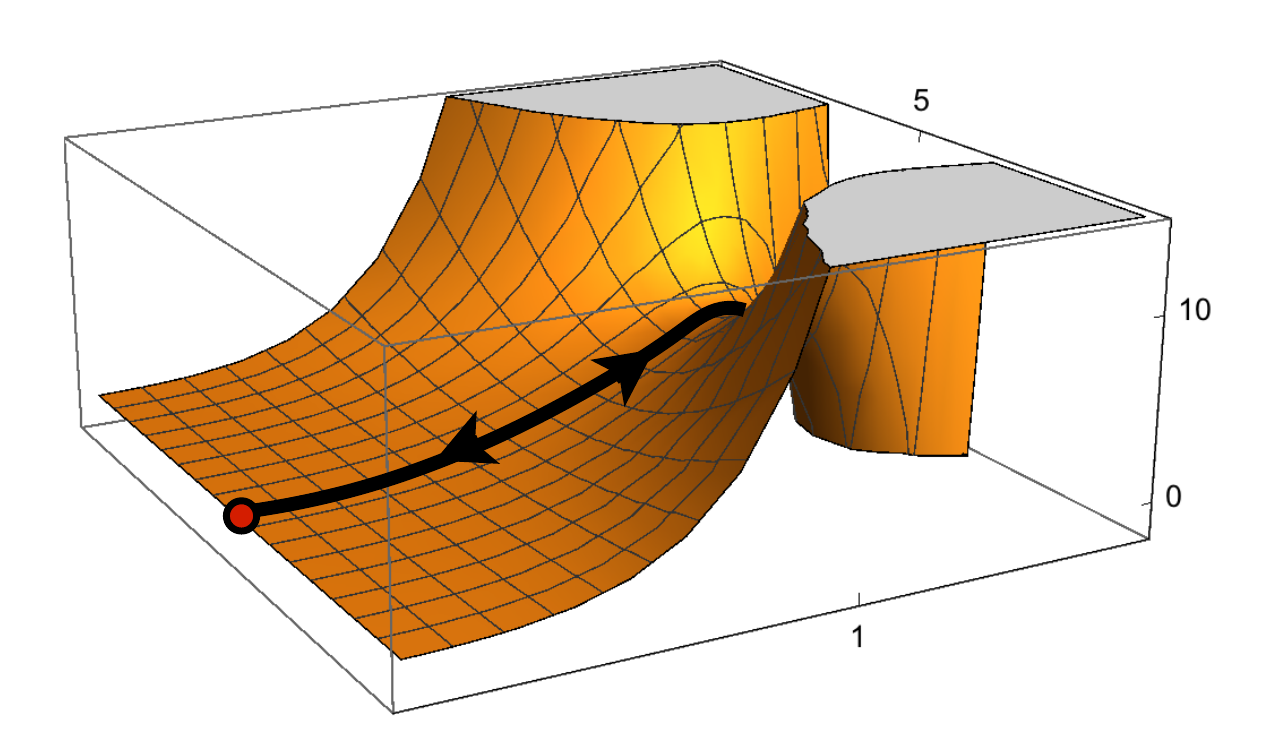}
	\caption{NS5 potential in the non-supersymmetric case, and path of the bounce.}
	\label{fig:ns5pot}
\end{figure}

The action for the NS5 in the massless region reads
\begin{align}
	S_{\mathrm{NS5}}&=-T \int d^6 \sigma e^{-2\phi}\sqrt{-g}|_{\Sigma_{NS5}}+T \int B_6 \nonumber\\
	&\label{eq:ns5-rho}
	=-T \mathrm{Vol}_{S^5}\int d\tau \sinh^5 \rho\left[e^{6A-2 \phi} \sqrt{\cosh^2 \rho + \dot \rho^2 + e^{-2A}g_{zz} \dot z^2}-\frac{h_7}6 \sinh\rho\right]\\
	&=-\frac{2^8 \pi T}{3^5 X^4 V_0^3}\int d\tau \sinh^5\rho\Big[
	(\dot\alpha^2-2X^5\alpha\ddot\alpha)\sqrt{\cosh^2\rho+\dot\rho^2+e^{-2A}g_{zz}\dot z^2}
	\nonumber
	\\
	&\hspace{4cm} -\frac{\sqrt{4X^5-1}}{\sqrt3} (\dot\alpha^2-2\alpha\ddot\alpha) \sinh\rho\Big ]
	\nonumber
\end{align}
where $T\equiv (2\pi l_s)^{-6}$. In fact, in the massless region, $\dot\alpha^2-2\alpha\ddot\alpha\equiv$ is constant (see eq. (\ref{eq:m0})). In both the supersymmetric and non-supersymmetric case, a solution can be obtained by taking $z$ to be such that $\dot\alpha=0$, which for our solution (\ref{eq:2D8s}) is at $z=\frac N2$. We can then compute $q$ by comparing with (\ref{eq:S-Ebrane}), (\ref{eq:q}):
\begin{equation}
	q_\mathrm{susy}= 1 \, ,\qquad q_\mathrm{non-susy}= \frac2{\sqrt3}>1\,.
\end{equation}
This indicates that the non-supersymmetric vacuum is unstable.

The physical potential, defined as $V= -S|_{\dot \rho= \dot z=0}$, is plotted in Fig.~\ref{fig:ns5pot}. The Euclidean potential is $-V$. We see that there is a local minimum at $z=N/2$, $\rho=0$ (red dot), a maximum for $\rho>0$, and that beyond that the potential becomes unstable for $\rho \to \infty$. The bounce solution starts at the local minimum, follows a trajectory in the $\rho$ direction past the maximum, and then comes back. It is best imagined as a ball rolling in the Euclidean potential $-V$.

While the quantum mechanics intuition of the previous approach is welcome, as we mentioned the computation is a lot simpler in the hybrid approach of (\ref{eq:hyb}), where we work in coordinates that make the bounce's spherical symmetry explicit. The NS5 wraps the $S^6$ in (\ref{eq:AdS-r}), and there is nothing to integrate: 
\begin{align}\label{eq:ns5-r}
	S_{\mathrm{NS5}}&=-T \mathrm{Vol}_{S^6}\left[e^{6A-2 \phi} \sinh^6r -h_7 \lambda\right]\\
	\nonumber
	&=-\frac{2^8 \pi T}{3^5 X^4 V_0^3}\Big[
	(\dot\alpha^2-2X^5\alpha\ddot\alpha)\sinh^6 r 
	-\frac{6\sqrt{4X^5-1}}{\sqrt3} (\dot\alpha^2-2\alpha\ddot\alpha) \lambda\Big ]
\end{align}
with $\lambda'= \sinh^6 r$. Extremizing the action with respect to $z$ and $r$ gives directly
\begin{equation}
	z= \frac N2 \, ,\qquad \tanh r = \frac{\sqrt 3}{2}
\end{equation}
in agreement with the previous results.

We can also see that the wall tension is related to the vacuum energy difference, as expected from the general comment around (\ref{eq:diff}). To see this quickly, evaluate the second term in either (\ref{eq:ns5-rho}) or (\ref{eq:ns5-r}); we see that at the point where $\dot\alpha=0$ it is proportional to $\alpha \ddot \alpha$, which for (\ref{eq:2D8s}) gives
\begin{equation}
	\frac{3^7}{2^3} n_0^2 \pi^4 \mu^2 (-3 N^2 + 4 \mu^2)\,.
\end{equation} 
This is also equal to $\partial_N a$, where $a= \frac{16}7 n_0^2 \mu^2 \left(N^3-4 N \mu^2 + \frac{16}5 \mu^3\right)$ \cite[(3.18)]{cremonesi-t} is the Weyl anomaly of the dual SCFT. This is in turn dual to the vacuum energy.

The bubbles we have found in this subsection reduce the value of $N$. This process can be iterated. As $N$ gets small, the supergravity approximation is less and less valid. Also, and perhaps earlier than that, at some point the two D8s will touch and the massless region will disappear. This will happen for $N=2\mu$ in (\ref{eq:2D8s}). Strictly speaking at this point our computation does not apply. However, formally it still does, if we restrict the motion of the NS5 to be at the midpoint $z=\frac N2$, which is now on top of the two now-coincident D8s.\footnote{To produce a figure like Fig.~\ref{fig:ns5pot} it would be necessary to also include the effect of D6-branes that get created when the NS5 ventures in the massive regions, but that is clearly more challenging, and we think it is not needed.} So all solutions that can be described in the supergravity approximation seem to be unstable, either perturbatively or non-perturbatively. 

Similarly for solutions with O6s, we expect that only the class with a single D8-brane on the left and on the right will survive the 7d perturbative stability analysis. In addition, since the massless region will differ only by the $S^2$ fiber being replaced by an $\mathbb{RP}^2$ \cite{afrt, apruzzi-fazzi}, an analogous NS5 bubble instability holds for these solutions. The cases with O8 require a more careful analysis, and we plan to come back to this in the future. 

% subsection ns5 (end)

\subsection{NS5--D6 bubbles for the smeared-D6 solution} % (fold)
\label{sub:bub-sm}

We now consider bubbles for the solution (\ref{eq:smD6}) with smeared D6-branes \cite{blaback-danielsson-junghans-vanriet-wrase-zagermann-1,blaback-danielsson-junghans-vanriet-wrase-zagermann-2}. 

Again we would like to consider NS5 bubbles; however, for this solution the $H$ flux $N$ is related to the number of D6-branes by $k=n_0 N$, where $F_0= \frac{n_0}{2\pi}$ is the Romans mass. So a change in the NS5-brane flux also implies a change in the number of D6-branes; in other words, $n_0$ D6-branes should end on the NS5 bubble. Another way to see this is to integrate the Bianchi identity (\ref{eq:bianchiF2sm}) on a sphere that surrounds the NS5.

There are two cases that we can consider: if the D6-branes are ending on the NS5 from outside, or from inside.  In other words, we can consider the number of D6-branes in the vacuum inside the bubble to have fewer or more D6-branes. We introduce a sign $\sigma$ which is $+1$ for an NS5, $-1$ for an anti-NS5. From the Bianchi identity $dH=-4\pi^2\sigma \delta$ we find that $N_\mathrm{out}- N_\mathrm{in}=\sigma$; the variation of D6-branes is then $k_\mathrm{out}- k_\mathrm{in}=\sigma n_0$. So if $\mathrm{sign}(\sigma n_0)=1$, we take the D6-branes to end from outside, and if $\mathrm{sign}(\sigma n_0)=-1$ we take the D6-branes to end from the inside.

The computation is now similar to (\ref{eq:ns5-r}), except that we now also include the D6s ending on the NS5. This gives
\begin{equation}
	S_{\text{tot}}=12^3 \frac{R^6}{g_s^2}\text{Vol}_{S^6} 
	\left( - \sinh^6 r_0 - \frac {5\sqrt{12}}{\sqrt7}\sigma \mathrm{sign}(F_0) \lambda(r_0) 
	-\sqrt{12} R g_s \sigma F_0 \int_{r_0}^{r_1} dr \sinh^6 r\right)\,,
\end{equation}
the last term being the D6 DBI contribution. $r_0$ is as usual the bubble location, whereas from the discussion above $r_1=\infty$ for $\mathrm{sign}(\sigma n_0)=1$ and $r_1=0$ for $\mathrm{sign}(\sigma n_0)=-1$. (Notice that the last term is negative in both cases.) In fact $\lambda$ was also defined below (\ref{eq:ns5-r}) as the integral of $\sinh^6 r$, so extremizing simply gives
\begin{equation}
	0= \sinh^5r_0 \left(-6 \mathrm{cosh}r_0 - \sqrt{\frac{12}7}\sinh r_0 \sigma \mathrm{sign}(F_0) (5 - 2)\right)\,.
\end{equation}
This gives $\tanh r_0 = - \sigma \mathrm{sign}(n_0) \sqrt{\frac 73}$, which does not have solutions. 

So, if we take into account that some D6s need to end on the NS5 instanton, they forbid the existence of this decay channel and the solution with smeared D6s is not destabilized.

% subsection ns5m (end)

% section bubbles (end)
 
\section{Domain walls} % (fold)
\label{sec:dw}

The bubbles we have considered so far are transverse to the radial coordinate in (\ref{eq:glo-ads}). We will now consider configurations which are transverse the Poincar\'e radial coordinate; we will call these domain walls. 

In section \ref{sub:glo} we will show that they too can be used as a window into stability, giving the same result as bubbles. Stable domain walls correspond to vacua of the dual SCFT, or equivalently to RG flows. In section \ref{sub:dw} we will give two examples of stable and BPS domain walls, and explain the dual SCFT interpretation. 

\subsection{Global and Poincar\'e coordinates} % (fold)
\label{sub:glo}

In Poincar\'e coordinates, AdS$_d$ reads
\begin{equation}\label{eq:poi-ads}
	 d \sigma^2 + e^{2 \sigma} ds^2_{\mathrm{Mink}_{d-1}} \,. 
\end{equation}
These coordinates are dual to studying a dual CFT in flat space, while the coordinates of (\ref{eq:glo-ads}) are dual to the CFT on $S^{d-1}$. 

A domain wall is a solution where $e^{2 \sigma}$ in (\ref{eq:poi-ads}) is replaced by a function $e^{2 a(\sigma)}$, such that as $\sigma \to \pm \infty$ the function is asymptotically linear, $a \sim a_\pm \sigma$. Such a solution preserves the $\mathrm{ISO}(d-2,1)$ symmetry that is manifest in (\ref{eq:poi-ads}), and interpolates between two copies of AdS$_d$ with cosmological costants related to $a_\pm$. This is holographically dual to a renormalization group (RG) flow relating two CFTs, with $\sigma\to \pm \infty$ corresponding to the ultra-violet (UV) and infra-red (IR) theories respectively. Depending on the behavior of the other fields at $\sigma\to + \infty$ (for example if the scalars behave in a normalizable or non-normalizable way) the dual RG flow is triggered by adding an operator ${\mathcal O}$ to the CFT, or by giving a vacuum expectation value (vev) to it. 

Here we are especially interested in domain walls that are well-localized in $\sigma$, so that the flow happens abruptly around a certain $\sigma= \sigma_0$. One such a domain wall is a brane, especially when the flux it sources is already present in large amounts in the AdS solutions at $\sigma \to \pm \infty$; in such a case its back-reaction can usually be neglected, or in other words it can be treated as a probe. The situation is similar to that in section \ref{sec:bubbles}, except for the locus of the brane.

Such localized domain walls have a normalizable behavior at $\sigma\to +\infty$, and are interpreted as vev RG flows. We can then consider them in turn as dual to points in the moduli space of the UV CFT. For example, in the canonical AdS$_5\times S^5$ example, we can consider D3-branes placed at $\sigma=\sigma_0$. These can be thought of as D3-branes that have been taken away from the stack of $N$ D3-branes that created the AdS solution by a near-horizon limit. The moduli space of such D3-branes is given by the value of $\sigma_0$, together with their position in the internal $S^5$. This is nothing else than $\mathbb{R}^6$, which can be thought of as the abelian moduli space of ${\mathcal N}=4$ super-Yang--Mills. The story is similar for more general AdS$_5$ solutions where the internal space is a Sasaki--Einstein manifold. 

So the moduli space of static branes is dual to the moduli space of vacua of the CFTs. We now wonder what happens with branes that are not static: namely, for those that sit at a $\sigma_0$ which is a function of the time coordinate in the Mink$_{d-1}$ part of (\ref{eq:poi-ads}). 

If such a brane feels a potential, it is natural to think that it is now signaling the presence of a potential $V_\mathrm{CFT}$ in the holographic dual. Such a $V_\mathrm{CFT}$ is heavily constrained by conformal invariance. Dual to this, the potential felt by a brane will be constrained to be a power law. The coefficient will receive a contribution from the gravitational potential that makes it shrink, and a contribution from the flux that makes it expand. If the latter wins, it is natural to imagine that it signals a potential in the dual CFT that has a wrong sign, in turn pointing to an instability. 

This was for example the point of view taken in \cite{gaiotto-t}. In that paper, a set of ${\mathcal N}=0$ supersymmetric solutions were found. However, in section 4.1.2 in that paper probe domain walls were considered; the result was that for D2-branes the flux contribution always wins. According to the point of view we just explained, this should signal an instability. 

We can show more explicitly that this test involving domain walls is in fact equivalent to the bubbles of section \ref{sec:bubbles}. The flux $F_d$ of section \ref{ssub:mms} is proportional to the volume form; in Poincar\'e coordinates (\ref{eq:poi-ads}) it reads
\begin{equation}
  F_d = f \,\text{vol}_{\text{AdS}_d} =f e^{(d-1)\sigma} d \sigma \wedge \text{vol}_{\text{Mink}_{d - 1}}\,.
\end{equation}
The potential can be chosen to be
\begin{equation}
  C_{d-1} = \frac{f e^{(d-1)\sigma}}{(d - 1)} 
  \text{vol}_{\text{Mink}_{d - 1}} \, .
\end{equation}
The action (\ref{eq:S-brane}) is now 
\begin{equation}
	S_{\text{Dp}} = \text{Vol}_{\text{Mink}_{d - 1}}e^{(d-1)\sigma}
   \left(-T  + \frac{Q f}{(d - 1)}\right) \,.
\end{equation}
The sign of the total coefficient depends again on whether $q$ in (\ref{eq:q}) is larger than 1, in which case the flux term wins and the brane is driven to the boundary of AdS; or if it is smaller than 1, in which case the gravitational term wins. So, if a brane initially at $\sigma=\sigma_0$ in Poincar\'e coordinates (\ref{eq:poi-ads}) tends to expand, so will a brane initially at $\rho=\rho_0$ in global coordinates.

% subsection glo (end)

\subsection{Domain walls for the supersymmetric solutions} % (fold)
\label{sub:dw}

We will now consider some domain walls in the AdS$_7$ supersymmetric solutions. They are BPS; as such they do not represent an instability, but rather supersymmetric RG flows. 

BPS branes are found most easily using calibrations. The relevant forms are the pure spinors used to find the solutions \cite[Sec.~4.5]{afrt} (correcting typos in that formula):
\begin{align}\label{eq:Psi}
	\Psi_0 & = \frac{1}{\sqrt{\dot \alpha^2 - 2 \alpha \ddot \alpha}}
	\left( i \dot \alpha 1_2 +   \sqrt{-2 \alpha \ddot \alpha} y^i \sigma_i\right) \\
  \Psi_1 & =  \frac{2^{1 / 4} \sqrt{\pi}}{\sqrt{\dot \alpha^2 - 2 \alpha \ddot \alpha}} \left( - \frac{\ddot\alpha}{\alpha} \right)^{1 / 4} 
\left( \sqrt{-2 \alpha \ddot \alpha} d  z
  1_2 + i (\dot\alpha d z y^i  + \alpha d y^i ) \sigma_i\right) \\
  \Psi_2 & =  - i \star_3 \Psi_1 \\
  \Psi_3 & =  - i \star_3 \Psi_0 
\end{align}

The calibration for space-filling branes is
\begin{equation}
  \frac12 e^{6 A - \phi} \mathrm{Tr} \left(\Psi_- + i e^A d\sigma \Psi_+\right)\,.
\end{equation}
where the trace is over the matrix indices in (\ref{eq:Psi}), and $_\pm$ denotes sum over all even or odd degrees. By definition, the pull-back of this form over the brane should be proportional to $e^{{\mathcal F}} \mathrm{vol}$, where as usual $\mathcal{F} = 2 \pi f + B$ and $\mathrm{vol}$ is the brane's internal volume form. Equivalently, the pull-back of another triplet of forms should vanish:
\begin{equation}
	\frac12 e^{6 A - \phi} e^{{\mathcal F}}\wedge\mathrm{Tr} \left[ \sigma_i\left(\Psi_- + i e^A d\sigma \Psi_+\right)\right]=0\,.
\end{equation}

\begin{figure}[ht]
	\centering
	\subfigure[\label{fig:RG-D8}]{\includegraphics[width=8cm]{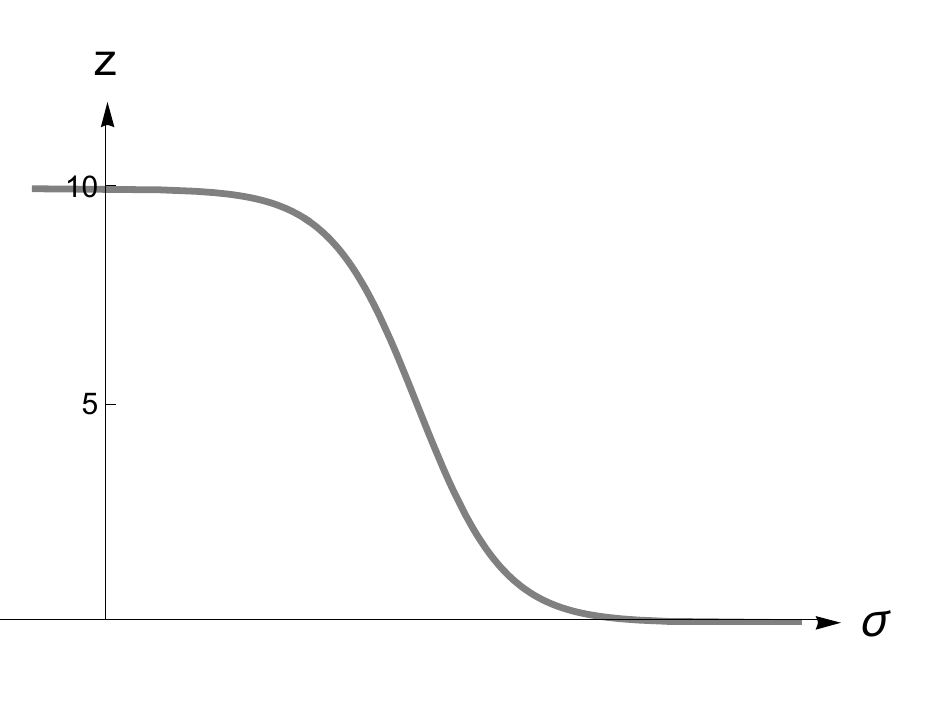}}\hspace{1cm}
	\subfigure[\label{fig:RG-D6}]{\includegraphics[width=6cm]{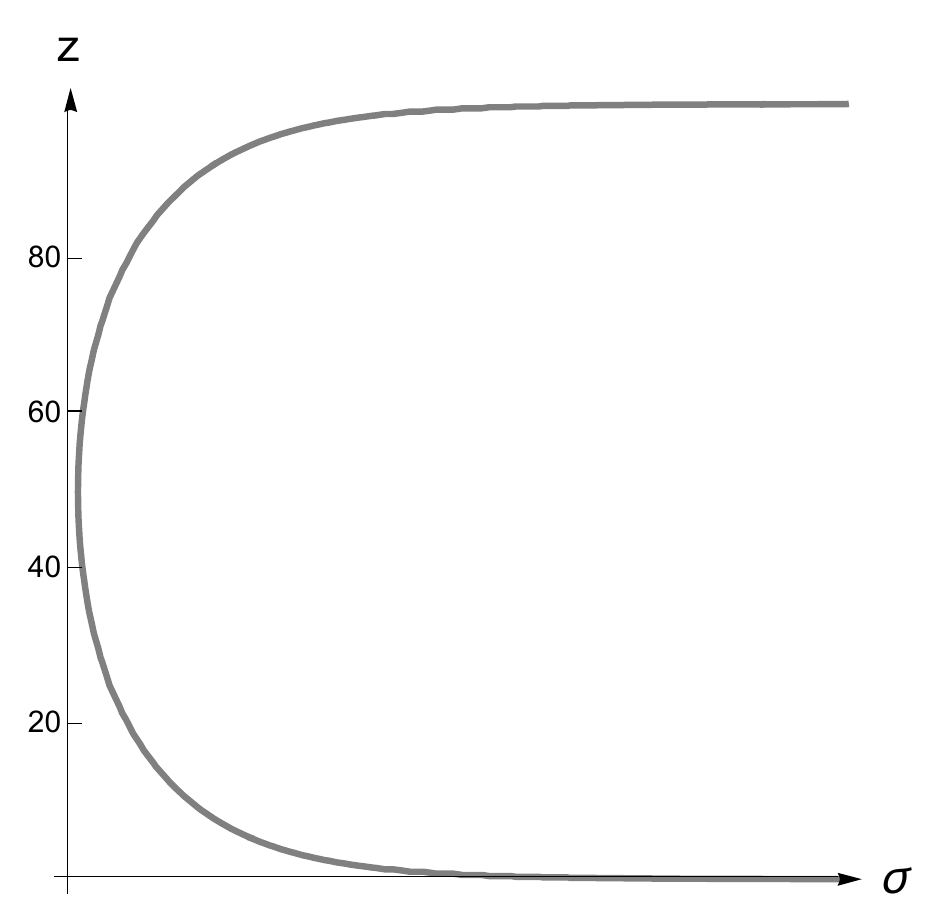}}
	\caption{Plots of the positions of a D8-brane and of a D6-brane domain wall.}
	\label{fig:RG}
\end{figure}

In the D8 case, the worldvolume is along the $S^2$, and is oblique in the $z$ and $\sigma$ directions. Taking $f=\frac k2 \mathrm{vol}_{S^2}$ we obtain the differential equation
\begin{equation}
  \partial_z \log (e^\sigma \alpha^{1 / 4}) = \frac{1}{4 (z - a)}
\end{equation}
which is solved by
\begin{equation}\label{eq:d8flow}
   z-a=c\, \alpha\, e^{4 \sigma} \,,
\end{equation}
with $c$ a constant. For $c=0$ this is not a domain wall at all: it is a D8 sitting at $z=a$ for any $\sigma$. This is the appropriate BPS locus for a background D8 with D6-charge equal to $a$, as we already observed in section \ref{sub:sol}. For $c\neq 0$, (\ref{eq:d8flow}) has the asymptotics
\begin{equation}
	\sigma\to - \infty \ \text{(IR): } z=a \, ,\qquad \sigma \to + \infty \ \text{(UV): } \alpha=0 \,.
\end{equation}
So in the UV the D8-brane tends to sit at a pole, where it shrinks into a stack of $k$ D6-branes. Which pole is chosen depends on the sign of $c$. In the IR it positions itself at $z=a$. We show a plot in Fig.~\ref{fig:RG-D8}, obtained for $a=10$ and with $\alpha$ in (\ref{eq:2D8s}).

This result is quite reasonable: it corresponds to a flow where the $\mathrm{SU}(a)$ symmetry of a stack of D6-branes, which is an $\mathrm{SU}(a)$ flavor symmetry in the dual SCFT$_6$, gets Higgsed, triggering an RG flow to a different SCFT$_6$. BPS flows of this type were discussed in \cite{deluca-gnecchi-lomonaco-t} from the point of view of gauged supergravity; here we see them from the point of view of D8 probes. 

We consider next the case of a D6-brane. Here the only difference is that the worldvolume does not include the $S^2$. The differential equation now reads
\begin{equation}
  \partial_z \log (e^\sigma \alpha^{1 / 4}) = 0 
\end{equation}
leading to 
\begin{equation}
	  e^{4 \sigma} = \frac c{\alpha}\,.
\end{equation}
In this case, for $\sigma\to + \infty$ (the UV) we have $\alpha\to0$, which has the two poles as a solution. As we decrease $\sigma$ towards the IR, there are still two solutions, which eventually join for some critical value of $\sigma$. We show this in Fig.~\ref{fig:RG-D6}.

This RG flow corresponds to a Higgsing that reduces the rank of all gauge group by one unit. From the point of view of the brane diagram, it corresponds to taking a D6 off.

% subsection dw (end)

% section dw (end)

\section*{Acknowledgements}

We would like to thank Giuseppe Dibitetto, Kate Eckerle, Luca Martucci and Eran Palti for interesting discussions. GBDL and AT are supported in part by INFN. AT is also supported by MIUR-PRIN contract 2017CC72MK003. GLM is supported by the Swedish Research Council grant number 2015-05333.

\bibliography{at}
\bibliographystyle{at}

\end{document}